\documentclass[12pt]{article}
\usepackage[utf8]{inputenc}
\usepackage[english]{babel}
\usepackage{hyperref}
\usepackage{paralist}
\usepackage{graphicx}
\usepackage{authblk}
\usepackage{natbib}
\usepackage{enumitem}
\usepackage{color}
\usepackage{textcomp}
\usepackage{booktabs}

\begin{document}


\title{Radiative-equilibrium model of Jupiter's atmosphere and application to estimating stratospheric circulations}

\author[1]{Sandrine Guerlet\thanks{Corresponding author: sandrine.guerlet@lmd.jussieu.fr}}
\author[1,2]{Aymeric Spiga}
\author[1]{Hugues Delattre}
\author[3]{Thierry Fouchet}

\affil[1]{\footnotesize 
LMD/IPSL, Sorbonne Université, ENS, PSL Université, École polytechnique,
Institut Polytechnique de Paris, CNRS, 
Paris, France 
\normalsize}
\affil[2]{\footnotesize 
Institut Universitaire de France,
Paris, France
\normalsize}

\affil[3]{\footnotesize 
LESIA, Observatoire de Paris, PSL Université, Sorbonne Université, Université de Paris, CNRS, 5 place Jules Janssen, 92195 Meudon, France
\normalsize}

\maketitle

\newpage
\section*{Abstract}

Jupiter's upper troposphere and stratosphere are host to a rich dynamical and chemical activity. This modulates the thermal structure and distribution of trace species and aerosols, which, in turn, impact the atmospheric radiative budget and dynamics.
In this paper, we present a computationally efficient 1-D seasonal radiative model, with convective adjustment, of Jupiter's atmosphere. Our model takes into account radiative forcings from the main hydrocarbons (methane, ethane, acetylene), ammonia, collision-induced absorption, several cloud and haze layers and an internal heat flux. 
We parametrize four tropospheric cloud and haze layers. Two of them (one tropospheric cloud near 800~mbar, one upper tropospheric haze, one stratospheric haze) are set to be uniform with latitude. On the contrary, we prescribe the spatial distribution of another UV-absorbing "polar" stratospheric haze comprising fractale aggregates based on published observational constraints, as their concentration varies significantly with latitude. 
We detail sensitivity studies of the equilibrium temperature profile to several parameters (hydrocarbon abundances, cloud particle sizes and optical depths, optical properties of the stratospheric polar haze, etc.). 
We then discuss the expected seasonal, vertical and meridional temperature variations in Jupiter's atmosphere and compare the modeled thermal structure to that derived from Cassini and ground-based thermal infrared observations.

We find that the equilibrium temperature in the 5--30~mbar pressure range is very sensitive to the chosen stratospheric haze optical properties, sizes and number of monomers. 
One of the three sets of optical properties tested yields equilibrium temperatures that match  well, to first order, the observed ones. In this scenario, the polar haze significantly  warms the lower stratosphere (10--30~mbar) by up to 20K at latitudes 45--60\textdegree , and reproduces an observed north-south asymmetry in stratospheric temperature. 
The polar haze also acts to shorten significantly the radiative timescales, estimated by our model to 100 (Earth) days at the 10-mbar level. 
At pressures lower than 3~mbar, our modeled temperatures  systematically underestimate the observed ones by $\sim$5K. 
This might suggest that other processes, such as dynamical heating by  wave breaking or by eddies, or a coupling with thermospheric circulation, play an important role. 
An alternate possibility is that the uncertainty on the abundance of hydrocarbons is responsible for this mismatch.
In the troposphere, we can only match the observed lack of meridional gradient of temperature by varying the internal heat flux with latitude.

We then exploit knowledge of heating and cooling rates (using our radiative seasonal model combined to observational constraints on the temperature) to diagnose the residual-mean circulation in Jupiter's stratosphere. This is done under the assumption that the eddy heat flux convergence term is negligible. 
In the Earth's stratosphere, the residual-mean circulation obtained with this method represents well, on a seasonal scale, the transport of tracers in regions where wave breaking and dissipation are weak.
However, on Jupiter, in the lower stratosphere (5--30 mbar), the residual-mean circulation strongly depends on the assumed properties of the stratospheric haze. 
Our main conclusion is that it is crucial to improve our  knowledge on the different radiative forcing terms (in particular regarding the stratospheric haze properties) to increase our confidence in the estimated circulation. By extension, this will also be crucial for future 3D GCM studies.

\section*{Highlights}

\begin{itemize}
    \item A seasonal radiative-convective model of Jupiter's atmosphere is presented.
    \item Stratospheric polar haze greatly impact the equilibrium temperatures.
    \item We evaluate the residual-mean stratospheric circulations and discuss caveats.
\end{itemize}


\newpage

\section{Introduction}
\label{intro}

Jupiter's troposphere hosts a rich dynamical activity with strong, alternately eastward and westward zonal jets at low and mid-latitudes, many vortices in the polar regions \citep[unveiled by Juno,][]{Adriani2018}, numerous planetary-scale and mesoscale waves, hotspots and disturbances \citep[e.g.,][]{Choi2013}. Jupiter's stratosphere is as dynamically active -- if not more -- than the troposphere, yet has received less attention comparatively to the large body of modeling work on the jovian tropospheric dynamics \citep{Will:03,Heim:05,Show:07,Schn:09,Young2019}. 
The observed temperature field features numerous wave signatures  \citep{Li2006, Fletcher2017} and isolated disturbances, both in the tropics and in the auroral regions \citep{Flasar2004, Sinclair2017}. Furthermore, a large variability of stratospheric temperature from one Earth year to another is observed at Jupiter's equator \citep{Fletcher2016}. This is associated with the quasi-quadriennal oscillation (QQO), a periodic oscillation in zonal wind and temperature thought to result from wave-mean zonal flow interactions \citep{Leovy1991,Orton1991,Flasar2004,Simon2007,Cosentino2017}. 

Stratospheric circulations are still poorly known, and are currently mostly deduced from the observation of anomalies in the distribution of trace species.
For instance, in the middle stratosphere (1-10~hPa), the observed meridional distributions of ethane (C$_2$H$_6$) and acetylene (C$_2$H$_2$) -- main by-products of the methane photochemistry -- are found to be at odds with the predictions of one-dimensional (1-D) neutral photochemical models.
While acetylene is maximum at low latitudes, following the yearly-averaged insolation as is expected from photochemistry, long-lived ethane increases towards the poles \citep[e.g.,][]{Nixon2007}. 
Other puzzling observations are related to molecules produced following comet Shoemaker-Levy 9 (SL 9) impact in Jupiter's atmosphere in 1994 and their subsequent migration \citep{Moreno2003, Griffith2004, Lellouch2006, Cavalie2017}. One of the most striking results is that, 6.5 years following the impact, HCN was found to be efficiently mixed from the impact site (44\textdegree S) to northern mid-latitudes while CO$_2$ was found to be greatly enhanced near the south pole \citep{Lellouch2006}.
In an attempt to explain the  observed opposite distributions of  C$_2$H$_6$ and C$_2$H$_2$, or HCN and CO$_2$, several models including parameterizations of meridional and vertical diffusion and advection have been proposed \citep[e.g.,][]{Hue2018, Lellouch2006}. Unfortunately, none could satisfactorily reproduce the observations. 
In short, there is currently no consistent picture of Jupiter's stratospheric circulations and how the distributions of trace species are impacted by those circulations.
The mechanism(s) forcing the aforementioned stratospheric circulations are also unknown, in particular regarding the role of wave activity in the troposphere and stratosphere -- by analogy with the Brewer-Dobson circulation in the Earth's stratosphere \citep{Butc:14} -- and that of radiative processes.
In this paper, we focus on the precise evaluation of radiative forcings with a 1-D radiative equilibrium model while the study of wave forcing is devoted to future work.

Understanding in detail the radiative forcings in Jupiter's atmosphere is also key to interpreting the observed thermal structure. At mid-latitudes and near the 10-mbar level, a 5 to 10~K temperature contrast is reported between the summer and winter hemispheres despite Jupiter's low obliquity (3\textdegree) \citep{Fletcher2016}.  
An explanation was proposed by \citet{Zhang2013b} who reported large radiative relaxation timescales near 10~mbar, which could lead to a seasonal lag in the atmosphere's response to seasonal forcing. However, their study only included forcing from gaseous compounds.
In a follow-up study, \citet{Zhang2015} highlighted the importance of radiative forcing by stratospheric aerosols of auroral origin, which were found to dominate the radiative heating at mid- and high- latitudes (instead of methane). However, the impact of including these terms on the temperature field, and its seasonal variations, was not studied.

Regarding the upper troposphere (100--500~hPa) and focusing on the zonal-mean temperature, the cloudy equatorial zone is found to be 2--4K colder than the warmer and clearer north and south equatorial belts, while there is little meridional temperature contrast at mid-latitudes (30\textdegree N--70\textdegree N and 30\textdegree S--70\textdegree S) \citep{Fletcher2016}.
These temperature variations are supposedly linked to tropospheric circulations, however, the radiative contribution from clouds and aerosols have not been studied quantitatively.

The aforementioned findings and open questions suggest that a complex interplay of dynamical and chemical activity takes place in Jupiter's middle atmosphere, modulating the thermal structure and the distribution of trace species and aerosols, which in turn impact the radiative budget and dynamics. 
All these observations and open questions motivate the development of a Global Circulation Model (GCM) for Jupiter extending to the upper stratosphere. Such a model would eventually take into account three-dimensional (3-D) dynamics, radiative forcings, photochemistry, cloud/aerosol microphysics and the couplings between them, including troposphere-stratosphere interactions. 
Several attempts have been made in this direction \citep{Yamazaki2004, Schn:09, Showman2019, Young2019} illustrating the modeling complexity and high computational cost necessary to address the questions opened by observations. 

Our goal is to obtain a Jupiter GCM capable of combining radiative transfer with high-resolution dynamics, akin to the approach we followed for Saturn's atmosphere \citep{Guerlet2014,Spiga2020}. In this paper, we focus on the efficient parametrization of a radiative-convective  model in Jupiter's upper troposphere and stratosphere. This model is to be later coupled with a hydrodynamical solver to build a Jupiter GCM capable of studying both tropospheric and stratospheric circulations. 
Apart from being part of a GCM, such a 1-D radiative-convective model can be used to compute radiative timescales \citep{Kuroda2014, Zhang2013b} and, when confronted with observations, can be a useful tool to diagnose whether to first order radiative processes govern or not the thermal structure of the atmosphere \citep[e.g.][]{Guerlet2014}. 
\citet{Kuroda2014} have developed such a radiative equilibrium model of Jupiter's stratosphere. They investigated the sensitivity of the equilibrium temperature profiles to the assumed hydrocarbon abundances and compared their results to a reference temperature profile obtained near the equator by Galileo. 
However, \citet{Kuroda2014} 's model  neglected  the radiative impact of  tropospheric and stratospheric aerosols that are expected to play an important role in heat absorption and redistribution. 
In this paper, we propose to refine the approach proposed by \citet{Kuroda2014} by including the missing radiative contributions and to extend the comparisons of our seasonal model to more recent observations. 

Finally, knowledge of heating and cooling rates (diabatic forcings) can also be exploited to diagnose the residual-mean circulation in the stratosphere, as was done by \citet{West1992} and \citet{Moreno1997}. This permits an estimate of the stratospheric circulation and transport of tracers without building a GCM to resolve the dynamics, under the limiting assumption that eddy heat flux is negligible compared to diabatic forcings. 
In theory, this circulation can then be exploited to re-visit the interpretation of observed distribution of trace species, as was done by \citet{Friedson1999} to address the spreading of dust following the impact of comet SL-9.
The studies by \citet{West1992} and \citet{Moreno1997} were based on observations from the Voyager epoch, and an update of this type of work based on Cassini-era observations and an up-to-date radiative transfer model is needed. This is especially critical because, while \citet{West1992} and \citet{Moreno1997} both agreed on the importance of including heating by stratospheric aerosols, the circulations they obtained differ both quantitatively and qualitatively.

In what follows, we describe a state-of-the art radiative-convective model for Jupiter's atmosphere (as part of a GCM to be presented in another paper), present comprehensive comparisons to recent temperature observations and exploit knowledge of the net radiative heating field to compute the residual-mean circulation in Jupiter's stratosphere.
Section~\ref{sec:model} describes our radiative transfer model for Jupiter's upper troposphere and stratosphere that includes up-to-date spectroscopic parameters, an internal heat flux, radiative effects of tropospheric clouds and aerosols as well as stratospheric aerosols comprising fractal aggregates. 
We present the resulting thermal structure and compare it with recently published ground-based and Cassini observations in Section~\ref{sec:temperature}. We then detail our methodology to compute the residual-mean circulation in Section~\ref{circu} and discuss these results in Section~\ref{discussion}.

\section{Jupiter radiative-convective model}
\label{sec:model}
\subsection{Overall description}

Our Jupiter radiative-convective model is adapted from its Saturn counterpart, described in detail in \citet{Guerlet2014}.
The two giant planets Jupiter and Saturn share many characteristics and, as a result, the main physical parametrizations are the same: a $k$-distribution model is used to compute gaseous opacities \citep{Goody1989}, the radiative transfer equations (including multiple scattering and Rayleigh scattering) are solved under a two-stream approximation, and a convective adjustment scheme relaxes  -- instantaneously -- the temperature profile towards the adiabatic lapse rate when unstable lapse rates are encountered. 
An internal heat flux, set to 7.48  W.m$^{-2}$ as determined by \citet{Li2018}, is also added as a radiative flux at the bottom of our model. 

Jupiter's diurnal cycle is neglected: a sensitivity test shows that the maximum amplitude of diurnal temperature variations is less than 0.1 K. Similarly, given the long radiative timescales in Jupiter's atmosphere, heating and cooling rates are computed -- and the temperature updated accordingly -- every 10 jovian days.
We take into account Jupiter's small obliquity (3.13\textdegree) and the moderate eccentricity of its orbit (0.048) that is expected to play a role in the seasonal cycle. Jupiter's perihelion occurs at a solar longitude of  L$_s$=57\textdegree, which is close to the summer solstice in the northern hemisphere (defined as L$_s$=90\textdegree, L$_s$ being the heliocentric longitude of Jupiter counted from the northern spring equinox).
Hence, northern summer is expected to be warmer than southern summer - at least in the stratosphere where radiative timescales are shorter than a season \citep{Kuroda2014}. 
If Jupiter's seasonal forcing was dominated by eccentricity rather than obliquity, one could even expect to get warmer temperatures in southern ``winter'' (Ls=90\textdegree) than during southern ``summer'' (Ls=270\textdegree).

Apart from the orbital and planetary parameters, the magnitude of the internal heat flux and the absence of opaque rings, the main differences between Saturn and Jupiter radiative models relate to the gaseous composition as well as cloud and haze properties, detailed below.

\subsection{Gaseous opacities and $k$-distribution model}

Our Jupiter radiative model takes into account gaseous opacity from the three main hydrocarbons: methane (CH$_4$), ethane (C$_2$H$_6$), acetylene (C$_2$H$_2$), along with collision-induced transitions by H$_2$-H$_2$ and H$_2$-He. Through their infrared emissions, these molecules are the major stratospheric coolants, while atmospheric heating is primarily due to absorption of visible and near-infrared solar radiation by methane and aerosols. Furthermore, we also take into account opacity from ammonia (NH$_3$) that was previously neglected in the Saturn model, as is justified in section~\ref{sec_gas}. 

 As line-by-line calculations of absorption coefficients are too time-consuming for the GCM runs we are aiming at, we use the correlated-$k$ method for the computation of the atmospheric transmission at each time step.
 Correlated-$k$ coefficients are pre-tabulated offline on a 2D temperature-pressure grid comprising twelve temperatures points from 70 to 400K and nine pressure levels from 10 to 10$^{-6}$ bar (one level every pressure decade, plus one level at 0.5 bar as ammonia varies rapidly with altitude in this region). 
To obtain these tables, high-resolution absorption coefficient spectra $k(\nu)$ are first computed using the KSPECTRUM line-by-line model \citep{Eymet2016} for a mixture of gases (CH$_4$, C$_2$H$_6$, C$_2$H$_2$, NH$_3$) at each point of the ($T,p$) grid. The gaseous abundance profiles used are detailed in section~\ref{sec_gas}.
A Voigt line shape is assumed except for CH$_4$, for which we use the far-wing line shape of \citet{Hartmann2002}, adapted to an H$_2$ atmosphere.
In a second step, we discretize these spectra in large bands (typically 100--300 cm$^{-1}$ wide) and use the KDISTRIBUTION code \citep{Eymet2016} to compute correlated-k coefficients $k(g)$ for each spectral band and each (T,p) value. We sample the cumulative probability $g$ with 8 Gauss integration points from 0 to 0.95 and another 8 points from 0.95 to 1.
The spectral discretization and the number of bandwidths is a compromise between accuracy (which increases when small spectral intervals are chosen) and computation time. After multiple tests, we have selected 20 bands in the thermal infrared (10--3200~cm$^{-1}$ or  3~$\mu$m -- 1 mm) and 25 bands in the visible and near-infrared (2000--33000~cm$^{-1}$ or 0.3--5~$\mu$m). 
When running a radiative-convective simulation, these $k$-coefficients are interpolated at each time step to the local temperature and to the pressure grid of the radiative transfer model. All radiative-convective simulations presented in this paper use a pressure grid consisting of 64 levels between 3 and 10$^{-6}$~bar.

\subsection{Updates on methane spectroscopy}

Spectroscopic line parameters are extracted from the HITRAN 2016 database \citep{Gordon2017}. However, the CH$_4$ linelist is known to be incomplete beyond 7,900 cm$^{-1}$; in particular, a methane absorption band at 1 $\mu$m is missing entirely, which could lead to an underestimation of the atmospheric heating rates. 
To fill this gap,  we complete the HITRAN 2016 methane linelist with a recent linelist based on \textit{ab initio} calculations \citep{Rey2016, Rey2018}. This list is available in the 0--12,000 cm$^{-1}$ range and contains position, energy and intensity for nearly 3.5 millions of transitions (assuming an intensity cut-off of 10$^{-28}$ cm/molecule), where the HITRAN 2016 database contains about 340,000 transitions.
In order to limit the computation time, and because the HITRAN 2016 methane database is thought to be reliable up to 7,900 cm$^{-1}$, we choose to combine the two linelists, using the spectroscopic parameters of  \citet{Rey2016} only beyond 7,900 cm$^{-1}$. 
Furthermore, we now include the transitions of the isotopologues $^{13}$CH$_4$ and CH$_3$D that were previously neglected by \citet{Guerlet2014}. The isotopic ratio $^{13}$C/$^{12}$C is set to the terrestrial value (0.011) in agreement with Galileo measurements \citep{Niemann1998}, and the  ratio CH$_3$D/CH$_4$ to 7.79$\times$10$^{-5}$ \citep{Lellouch2001}. 
The spectroscopic line data of $^{13}$CH$_4$ and CH$_3$D, for the  spectral domain 0--12,000 cm$^{-1}$, also come from  \textit{ab initio} calculations by \citet{Rey2016}, which are more exhaustive than HITRAN 2016.
 
\begin{figure}
  \centering
  \includegraphics[width=\linewidth]{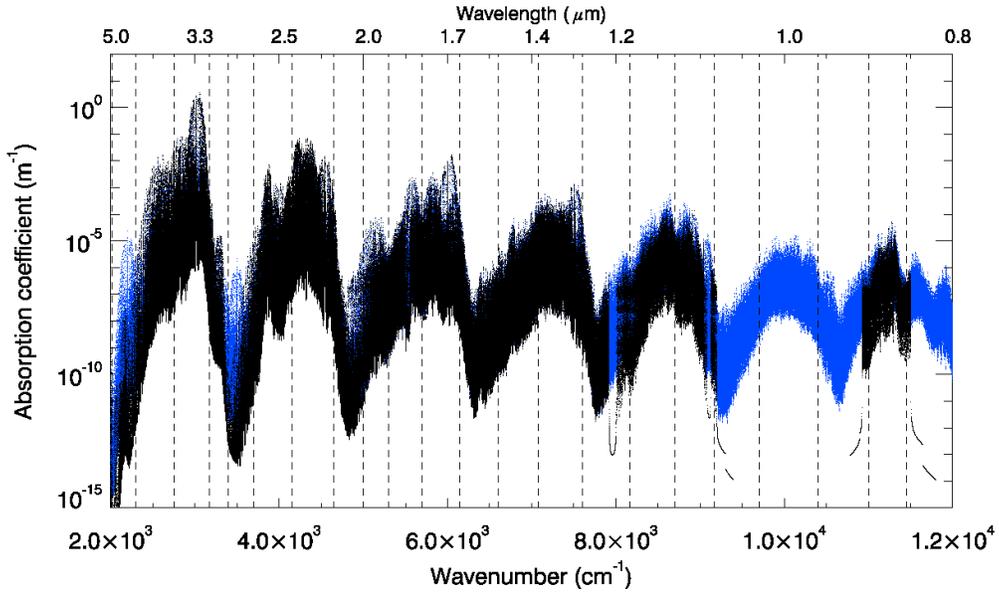}
  \caption{Absorption coefficient spectrum calculated for a pressure of 10~mbar and a temperature of 160K in the visible range from the HITRAN 2016 database (main $^{12}$CH$_4$ isotope, in black), and also including \citet{Rey2018} linelists for $^{12}$CH$_4$ (beyond 7,900 cm$^{-1}$), CH$_3$D and $^{13}$CH$_4$ (in the range 0--12,000 cm$^{-1}$), in blue. The vertical dashed lines represent the limits of the 22 bands used for generating $k$-distribution coefficients.}
  \label{fig:hitran}
\end{figure}

Figure \ref{fig:hitran} shows the comparison between absorption coefficient spectra in the visible range computed using the HITRAN 2016 database (considering $^{12}$CH$_4$ only) with the new combination of the HITRAN 2016 and \citet{Rey2018} linelists for $^{12}$CH$_4$, $^{13}$CH$_4$ and CH$_3$D. This figure illustrates the important addition of the  $^{12}$CH$_4$ \citet{Rey2018} linelist to HITRAN 2016 beyond 7,900~cm$^{-1}$, as well as the contribution of CH$_3$D that features emission bands at 1,100  cm$^{-1}$ (not shown), 2,200~cm$^{-1}$ and 3,500~cm$^{-1}$.  $^{13}$CH$_4$ lines are not visible in this figure as their main absorption bands are mingled with  $^{12}$CH$_4$.
Regarding the impact on the equilibrium temperature profile, using the $^{12}$CH$_4$ linelist from \citet{Rey2018} beyond 7,900~cm$^{-1}$ increases the heating rates by 10\% to 20\% compared to using HITRAN 2016 alone in the range 0--12,000~cm$^{-1}$. This corresponds to a stratospheric warming between 2 and 3.5~K, the maximum lying near 10-20 mbar. The addition of the two methane isotopologues yields a modest increase of $\sim$1~K.

We recall that for the Saturn model, \citet{Guerlet2014} used a combination of HITRAN 2012 (similar to HITRAN 2016 as far as CH$_4$ is concerned) up to 7,800~cm$^{-1}$ with another set of $k$-distribution coefficients computed in the range 7,800--25,000~cm$^{-1}$ based on the \citet{Karkoschka2010} methane band model. 
In the \citet{Guerlet2014} study, we concluded that the amount of heating by methane beyond 7,800~cm$^{-1}$ was significant, but we did not distinguish between the near infrared part (7,800-12,000 cm$^{-1}$) and the visible part beyond 12,000 cm$^{-1}$. 
In order to complete our study and evaluate the radiative heating resulting from absorption of visible solar photons, we computed a new set of $k$-distribution coefficients in the range 12,000--25,000 cm$^{-1}$ based on the  \citet{Karkoschka2010} data. We find that, as far as our Jupiter model is concerned, absorption by methane in this range has a negligible impact, warming the atmosphere by at most 0.4K at the 10-20 mbar level. This can be explained by the small volume mixing ratio of methane combined with its rapidly decreasing absorption coefficients beyond 12,000 cm$^{-1}$.

Hence, in this paper we choose to work only with the HITRAN 2016 and \citet{Rey2018} linelists and neglect gaseous absorption in the visible part (that is, beyond 12,000 cm$^{-1}$).  The near infrared part of the spectrum is discretized in 22 spectral intervals, shown in Figure \ref{fig:hitran}, to which we add three bands covering the visible part with zero gaseous opacity. These bands are needed to contain cloud and aerosol opacity.

\subsection{Gaseous abundances} 
\label{sec_gas}

In the present study, we neglect meridional variations of the trace species. Instead, the $k$-tables are computed for a single volume mixing ratio vertical profile for each species, reflecting Jupiter's average composition.
We set the H$_2$ volume mixing ratio to 0.863, the helium mixing ratio to 0.136 \citep{Niemann1998}, and the mixing ratio of CH$_4$  in the deep troposphere to the value determined in situ by the Galileo probe mass spectrometer (2.07~$\pm 0.5 \times$ 10$^{-3}$, \citet{Wong2004b}). We note that other values of the methane mixing ratio have been reported from independent, remote sensing measurements, such as \citet{Gautier1982}. They determined a value of 1.8~$\pm 0.2 \times$ 10$^{-3}$, which is consistent with the value of \citet{Wong2004b}, within error bars. 

The volume mixing ratio of CH$_4$ decreases with altitude primarily due to molecular diffusion in the upper stratosphere, and to a lesser extent due to photo-dissociation by solar UV radiation \citep{Gladstone1996,Moses2005}.
The altitude level of the methane homopause on Jupiter is estimated to lie in the range 10$^{-5}$ to 10$^{-6}$~bar based on stellar occultations \citep{Festou1981, Yelle1996,Greathouse2010}. This is significantly deeper than on Saturn, where the homopause level is estimated to a few 10$^{-7}$ bar. This difference is explained by a much stronger eddy mixing coefficient on Saturn compared to Jupiter \citep{Moses2005}. 
However, the exact homopause level on Jupiter is not well constrained by observations and differs among studies \citep{Greathouse2010}; it could also vary with time and latitude.
 Similarly, uncertainties on the eddy mixing coefficient and photodissociation rates map into a family of the methane vertical abundance profile in Jupiter photochemical models (see for instance models A, B and C of \citet{Moses2005}). 

\begin{figure}
  \centering
  \includegraphics[width=0.7\linewidth]{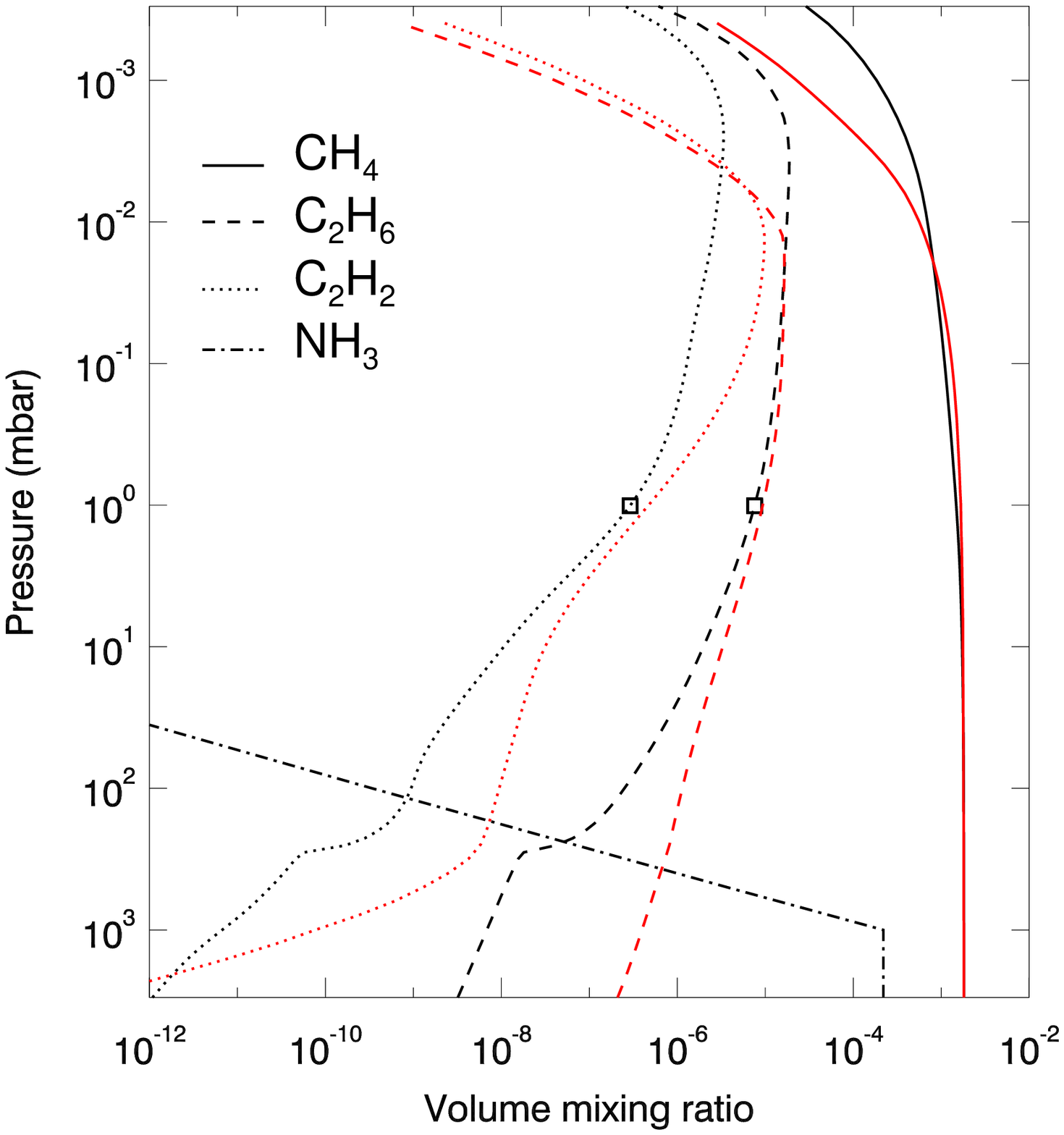}
  \caption{Vertical profiles for the volume mixing ratio of methane, ethane, acetylene and ammonia corresponding to an average of photochemical models ``A'' and ``C'' of \citet{Moses2005} (in black) or to the photochemical model used by \citet{Nixon2007} (in red). The  C$_2$H$_6$ and C$_2$H$_2$ vertical profiles of Moses et al. are scaled to the abundances retrieved by \citet{Nixon2010} at 1~mbar  and averaged between 40\textdegree S and 40\textdegree N (shown as squares).}
  \label{fig:hydro_profiles}
\end{figure}

The choice of a vertical profile for methane (and other hydrocarbons) is thus partly arbitrary and will influence the vertical energy deposition,  hence the resulting equilibrium temperature profile in the stratosphere, as was already reported by \citet{Zhang2013b} and \citet{Kuroda2014}. 
We choose to set the volume mixing ratio of the three hydrocarbons to the average of the 1-D photochemical models A and C  from \citet{Moses2005}. This corresponds to an homopause level at $\sim$1 $\mu$bar. Regarding C$_2$H$_6$ and C$_2$H$_2$, we further scale these model profiles so that the hydrocarbon abundances at 1~mbar match the low to mid-latitude Cassini/CIRS observations of \citet{Nixon2010}: 7.6 $\times$ 10$^{-6}$ for  C$_2$H$_6$, 2.9 $\times$ 10$^{-7}$ for C$_2$H$_2$. 
For the purpose of sensitivity tests, we also compiled a different set of $k$-tables with the hydrocarbon profiles set to the photochemical model profiles used by \citet{Nixon2007}, which feature a deeper homopause level ($\sim$10 $\mu$bar).
The different hydrocarbon vertical profiles are illustrated in Fig.~\ref{fig:hydro_profiles}.

\begin{figure}
  \centering
  \includegraphics[width=0.45\linewidth]{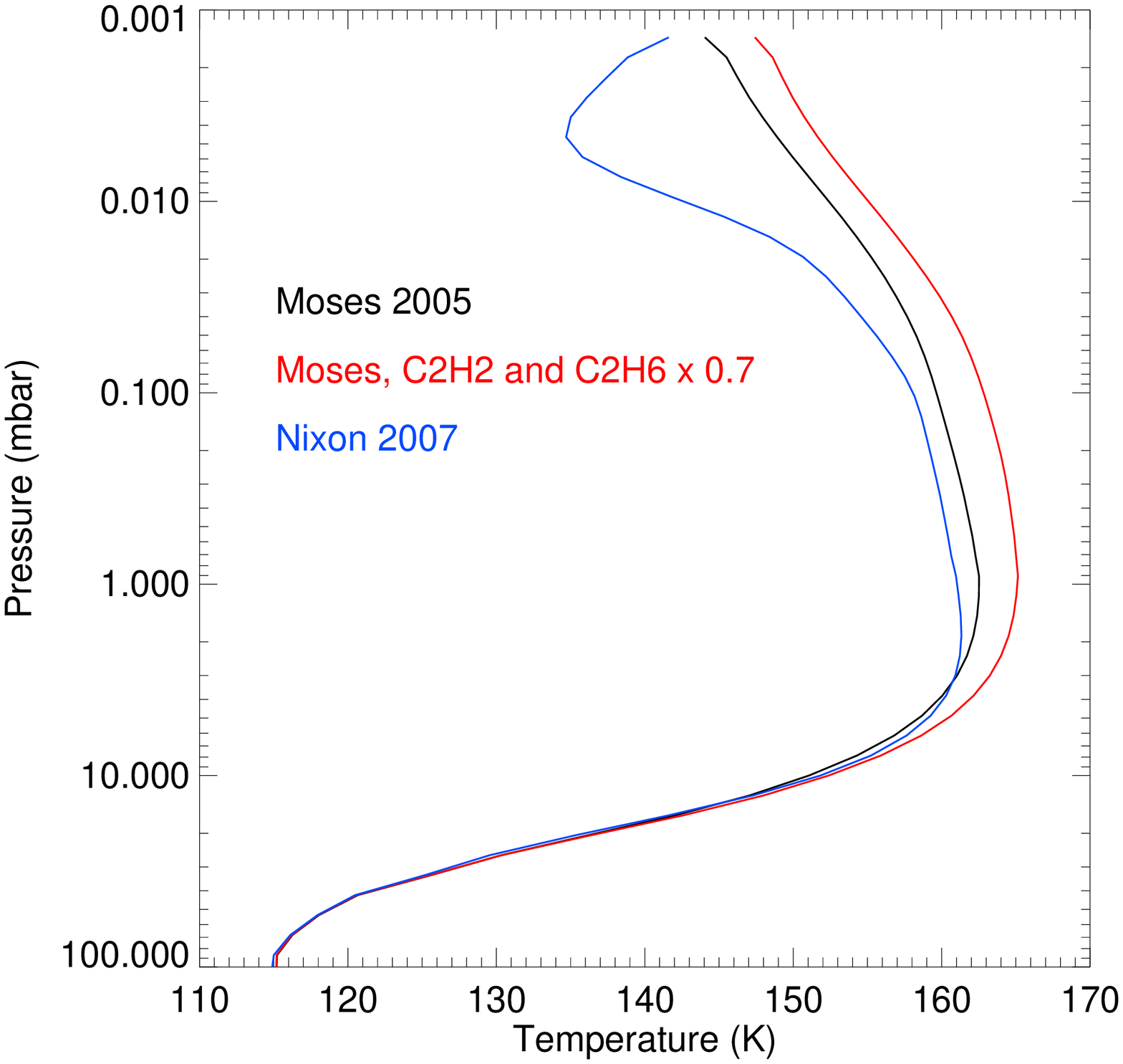}
 \includegraphics[width=0.45\linewidth]{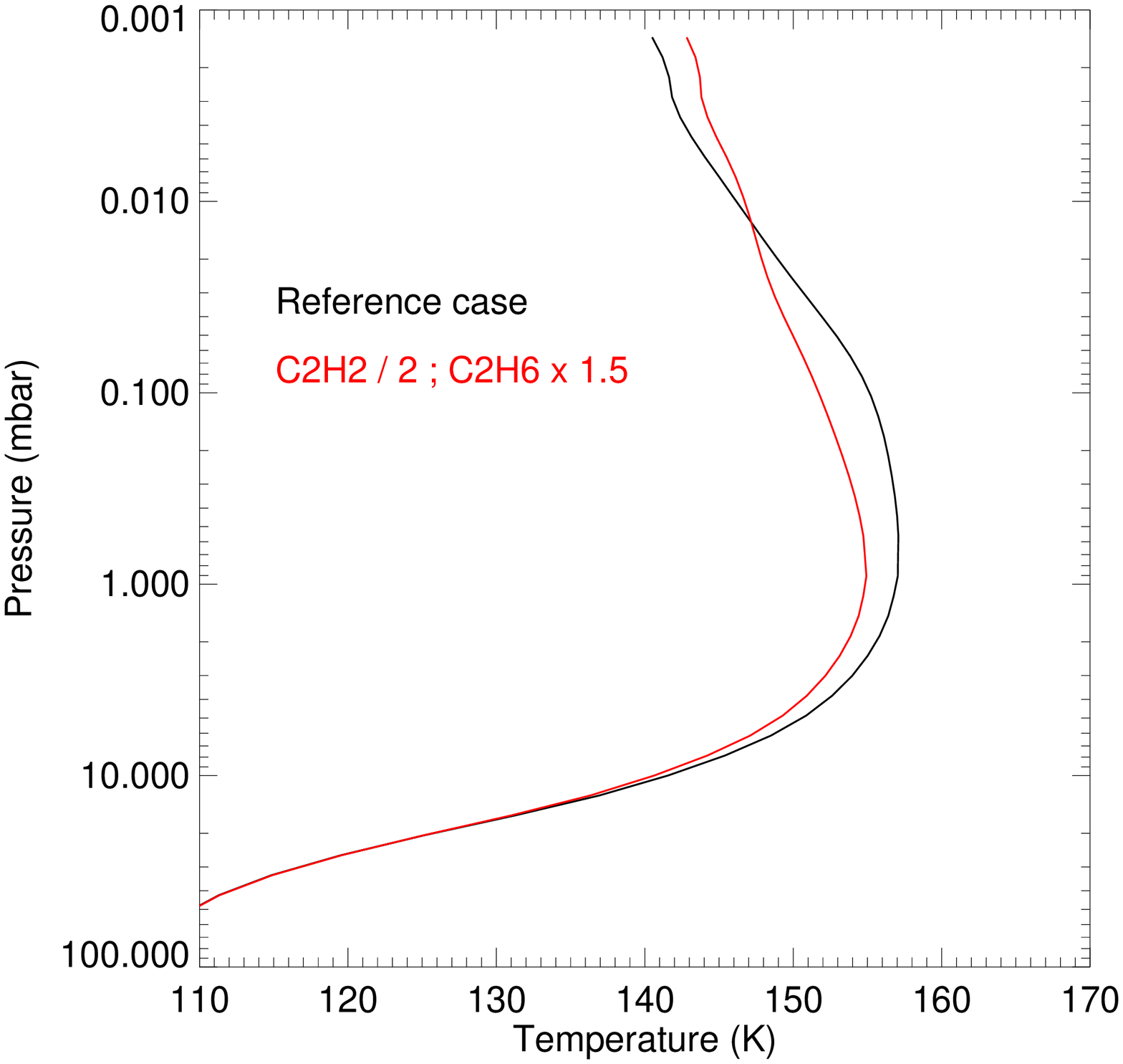}
  \caption{Equilibrium temperature profiles for different hydrocarbon mixing ratio profiles. Left: example at latitude 20\textdegree N, Ls= 0, with the hydrocarbon abundances set to that of \citet{Nixon2007} (in blue) or to the average of model A and C of \citet{Moses2005} (in black) or the latter but with 30\% less C$_2$H$_2$ and C$_2$H$_6$ (in red). Right: example at latitude 60\textdegree N, Ls= 0, with the reference hydrocarbon abundances \citep[the average of model A and C of][]{Moses2005} (black line) or with a 50\% increase in C$_2$H$_6$ and a 50\% decrease in C$_2$H$_2$ (red line).}
  \label{fig:tempe_hydro}
\end{figure}

We present in Fig.~\ref{fig:tempe_hydro} the impact of assuming different hydrocarbon profiles on the equilibrium temperature, based on aerosol-free 1-D radiative-convection simulation for latitude 20\textdegree N. In the 1~mbar to 10~$\mu$bar region, the \citet{Nixon2007} photochemical model has $\sim$1.5 to 3 times more acetylene than our combination of the \citet{Moses2005} models (but similar amounts of ethane and methane), resulting in greater cooling rates and stratospheric temperatures 2 to 5K colder. Between 1 and 10 $\mu$bar, the temperature calculated using the \citet{Nixon2007} hydrocarbons reaches a minimum, then increases with height. In this pressure range, all three hydrocarbons of the \citet{Nixon2007} model sharply decreases with altitude. This yields lower heating rates through lower absorption by CH$_4$ (explaining the cold temperatures near 5~$\mu$bar) but also lower cooling rates by hydrocarbon infrared emissions. As C$_2$H$_6$ and C$_2$H$_2$ decrease more sharply than CH$_4$, the net effect is a warming of the atmosphere between 5 and 1~$\mu$bar.

We also evaluate the impact, on our equilibrium temperature profile, to a 30\% decrease in both C$_2$H$_6$ and C$_2$H$_2$ mixing ratios with respect to our nominal hydrocarbon profiles based on the \citet{Moses2005} models. This 30\% change reflects typical observed meridional and temporal variations at low to mid-latitudes \citep{Melin2018}. This yields a temperature increase of about 3K above the 10-mbar pressure level.
This is in qualitative agreement with the work of \citet{Kuroda2014} who estimated a temperature change of $\pm$ 8K when C$_2$H$_6$ and C$_2$H$_2$ were divided or multiplied by two. 

Finally, we also quantify the impact of an increase of +50\% in ethane mixing ratio, while acetylene is divided by two: this case study corresponds to what is observed at high latitudes compared to the equator \citep{Nixon2010, Fletcher2016}. In doing so, we evaluate the impact of neglecting actual meridional variations, in the (realistic) case where acetylene and ethane exhibit opposite trends. We find that the impact of increasing ethane while decreasing acetylene is rather small, as there is a partial compensation of the two competing effects (an increase in radiative cooling rates when ethane is increased, a decrease of it when acetylene is decreased). A similar conclusion was reached by \citet{Zhang2013b}, as far as the Cassini/CIRS hydrocarbon retrievals were concerned. At 60\textdegree N and between 1 and 10~mbar, the resulting temperature profile  (shown in Fig.~\ref{fig:tempe_hydro}) is 1 to 2~K colder than the nominal case, and is up to 4~K colder in the 1 to 0.05~mbar range. This is because ethane is a more efficient coolant than acetylene is in this pressure range. The two temperature profiles are then similar at and around 0.01~mbar. At even lower pressures, the temperature becomes slightly warmer than the nominal case. This is because acetylene cools more efficiently the upper stratosphere than ethane, as was already mentioned by \citet{Kuroda2014}, so that a decrease by a factor of 2 of acetylene results in a net warming compared to the nominal case.
Hence, we conclude that the impact of neglecting meridional variations in ethane and acetylene on stratospheric temperatures is of the order of 2--4~K.

\begin{figure}
  \centering
  \includegraphics[width=\linewidth]{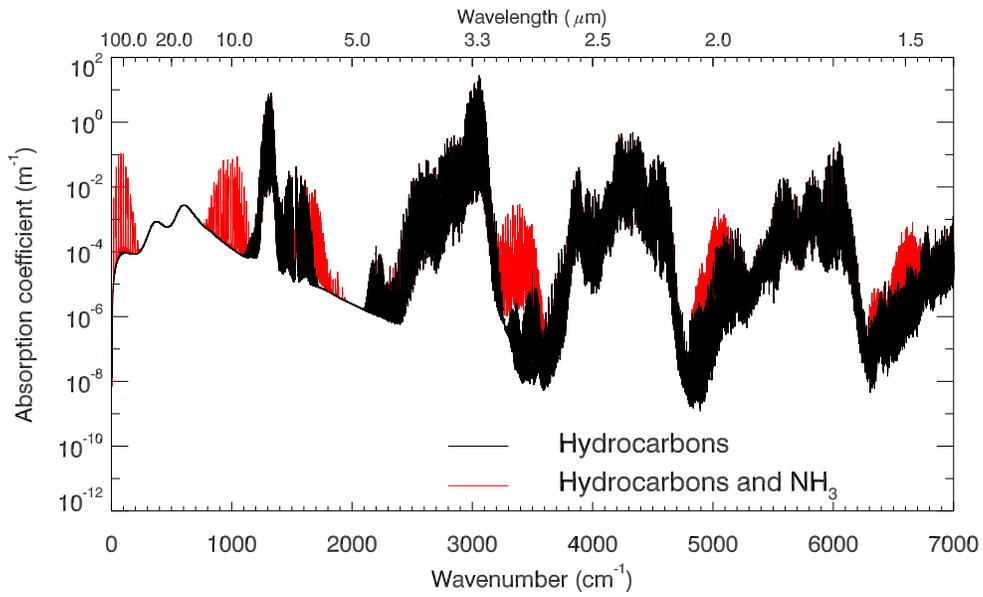}
  \caption{Absorption coefficient spectrum calculated for a pressure of 500~mbar and a temperature of 130K, including (in red) or not (in black) ammonia. Collision-induced absorption by H$_2$-H$_2$ and H$_2$-He is included and is important in the 5--100~$\mu$m range.}
  \label{fig:nh3_spectra}
\end{figure}

In addition to hydrocarbons, we evaluate the influence of including ammonia (NH$_3$). For the tropospheric temperatures encountered on Jupiter, ammonia is expected to condense at $\sim$0.7 bar ($\sim$150K) \citep{Atreya1999}. Following the vapour pressure curve, its mixing ratio rapidly decreases above this pressure level to become insignificant at tropopause levels. 
We set the ammonia "deep" mixing ratio (at 3 bar) to 250 ppm consistently with planet-average abundances measured by Juno at this pressure level \citep{Li2017} and assume a fractional scale height of 0.15 above the 0.7 bar level \citep{Nixon2007}. 
We find that including NH$_3$ in our model yields a significant temperature increase of 10K in the troposphere. This temperature increase is caused by the absorption of near infrared solar light by NH$_3$ and also by a small greenhouse effect. Indeed, adding ammonia increases the infrared opacity, especially beyond 5 $\mu$m, as shown in Fig.~\ref{fig:nh3_spectra}.
In consequence, thermal radiation emitted deep in the troposphere at long wavelengths is partly absorbed by ammonia in the mid-troposphere, which limits the cooling-to-space and warms the troposphere. 
We note that we have also tested including phosphine (PH$_3$) with a deep abundance of 6.0$\times$10$^{-7}$ and a fractional scale height of 0.3 \citep{Nixon2007}, but found a negligible impact on the thermal structure.

We choose here to keep the ammonia mole fraction constant with latitude. However, recent  measurements made by the Juno microwave radiometer revealed highly variable ammonia concentrations, hinting at an ammonia-rich equatorial region (300--340 ppm in the 1--3 bar pressure range) and an ammonia-depleted region at 10--20\textdegree N (as low as 140 ppm at 1 bar) \citep{Li2017}. 
A sensitivity test where NH$_3$ is decreased by 40\% (150~ppm at the 1-bar level instead of 250~ppm) yields a small temperature decrease of 1K in the troposphere.
Hence, the spatial variations derived from Juno should not significantly impact the thermal structure (in terms of direct radiative forcing). Having included ammonia in our Jupiter model does not challenge our previously published results on Saturn's thermal structure \citep{Guerlet2014}. Indeed, the upper tropospheric volume mixing ratio of NH$_3$ is 10 to 100 times lower on Saturn than on Jupiter, due to a deeper condensation level ($\sim$~1.4 bar), and we find that including NH$_3$ does not have a significant impact on Saturn's upper tropospheric temperatures.

\subsection{Treatment of tropospheric clouds and aerosols \label{sec:cloudaerosol}}

Cloud and haze particles are expected to play a key role in the radiative budget of Jupiter's troposphere. Through their vertical distribution, microphysical and optical properties, they control the local absorption of solar radiation at different depth, hence the temperature and heat redistribution.
Many studies have attempted to characterize their physical and chemical properties from remote sensing measurements, with more or less agreement between them due to the complexity of such ill-posed inverse problems, with non-unique solutions. A complete review on the cloud and haze observational constraints would be beyond the scope of this paper; instead we summarize below the main findings of the cloud and haze radiative impact in the upper troposphere relevant to our study, at pressures less than 2--3 bar.

\subsubsection{Observational constraints}

There is an overall consensus that, in order to reproduce both visible and thermal infrared imaging data, a combination of a diffuse haze comprising small particles (0.3--2~$\mu$m) located above a compact cloud comprising larger particles (3--100 $\mu$m) is needed (\textit{e.g.} \citet{West1986} from Pioneer data, \citet{Irwin2001} using Galileo/NIMS spectra, \citet{Wong2004a} using Cassini/CIRS data, \citet{Sromovsky2018} using New Horizon/LEISA).

The location of the cloud deck  is estimated to lie in the range 0.5--1.2~bar depending on the studies, while the upper tropospheric haze likely extends up to 150--300 mbar, \textit{i.e.} near the tropopause. Thermochemical equilibrium models predict that ammonia condenses at $\sim$700 mbar, while ammonium hydrosulfide (NH$_4$SH) is expected to form another cloud layer at $\sim$2 bar \citep{Atreya1999}. However, the spectroscopic signatures of NH$_3$ ice at 2~$\mu$m, 9.4~$\mu$m and 26~$\mu$m have been rarely observed, and \citet{Baines2002} showed that spectrally identifiable ammonia clouds cover less than 1\% of Jupiter's globe. Rather, \citet{Sromovsky2010b} suggest that the haze layer consists of small ammonia-coated particles overlying a cloud layer of NH$_4$SH ice particles  at $\sim$600 mbar, or that several layers of NH$_3$ and NH$_4$SH ice particles coexist, which would explain the lack of strong NH$_3$ absorption features in the infrared.
A similar conclusion was reached by \citet{Giles2015}, who used   Cassini/VIMS data between 4.5 and 5.2~$\mu$m to constrain Jupiter's cloud structure.
The authors find that VIMS observations can be modeled using a compact, highly reflecting cloud layer located at a pressure of 1.2 bar or lower, with spectrally flat optical properties in this spectral range. Indeed, setting the refractive index to that of pure NH$_3$ or NH$_4$SH ice particles could not fit VIMS observations, for any particle sizes in the range 1--40 $\mu$m. 

A few observational constraints exist on haze and cloud particles optical properties: Pioneer observations analyzed by \citet{Tomasko1978} require highly reflecting particles at 0.44 and 0.6~$\mu$m, with single scattering albedo higher than 0.95 for the haze, and higher than 0.98 for the cloud particles.
Typical cumulative optical depths  measured in the visible (0.75 $\mu$m) vary from 1 to 5  above the 500-mbar level (that can reasonably be attributed to the haze opacity), and vary between 5 and 20 above the 1-bar level \citep[see][and references therein]{Sromovsky2010b}. 
In the near infrared, haze optical depths varying between 0.5 and 5 have been derived at 2~$\mu$m \citep{Irwin2001, Kedziora2011}.
The optical depth variations between cloudy zones and less opaque belts likely stem from cloud optical depth variations \citep[found to lie between 8 and 22 at 5 $\mu$m,][]{Giles2015} rather than variations of the haze itself.

\subsubsection{Cloud model and sensitivity studies \label{sec:cloud_sensi}}

\begin{figure}
  \centering
  \includegraphics[width=0.8\linewidth]{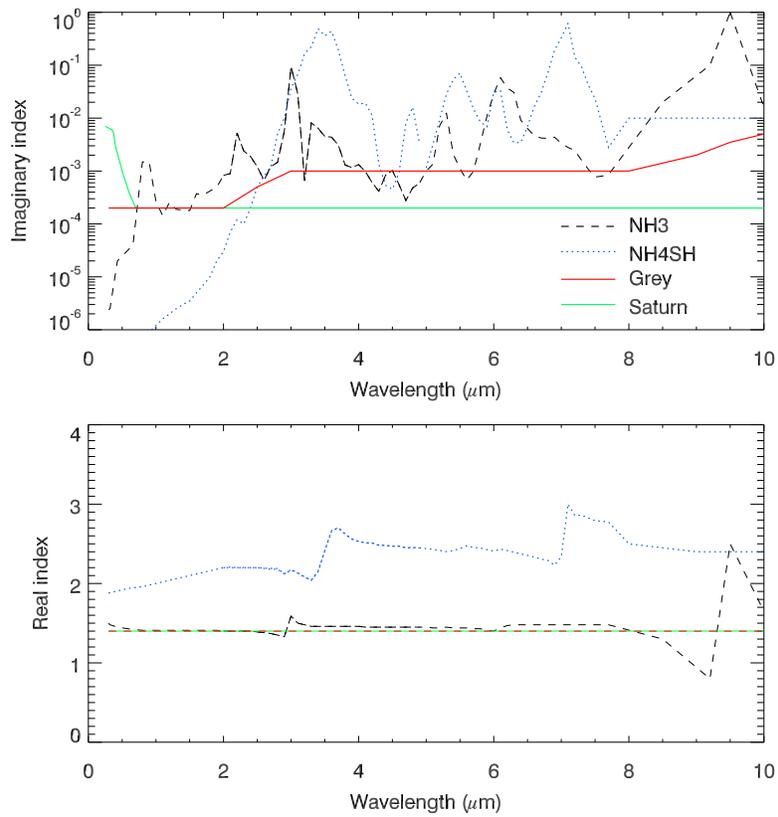}
  \caption{Imaginary (top) and real (bottom) refractive indexes of four assumed cloud or haze compositions. The ``Saturn'' real refractive index is the same as for the ``grey'' particle type.}
  \label{fig:refractive_indexes}
\end{figure}

Our goal here is to set up an effective cloud and haze model that would reproduce Jupiter's albedo, thermal structure and be consistent with the observed visible and infrared cloud optical depths and physical properties at a global scale.
We emphasize that this effective model is not meant for comparisons to detailed spectroscopic observations, but rather is meant to account for the radiative forcing of cloud and haze particles and their role in the radiative budget.
In what follows, we assume a two-layer cloud structure with an extended, upper haze located above a compact cloud and test the sensitivity to varying the cloud and haze composition (optical constants), optical depth, particle sizes and the altitude of the cloud deck.
We assume spherical particles and compute the optical properties (single scattering albedo, extinction coefficient and asymmetry factor) with a Mie scattering code.
Four compositions are tested: 
\begin{enumerate}
\item\label{comp:ammonia} pure NH$_3$ ice particles, with optical constants from \citet{Martonchik1984}; 
\item\label{comp:hydrosulf} pure NH$_4$SH ice particles, with optical constants from \citet{Howett2007}; 
\item\label{comp:saturn} same material as our Saturn haze model (\citet{Guerlet2014}, based on observational constraints from \citet{Karko1993} on Saturn); 
\item\label{comp:grey} particles with nearly grey optical constants. 
\end{enumerate}
Composition~\ref{comp:grey} has real and imaginary indices set arbitrarily close to that of NH$_3$ except for smoother spectral features (since the sharp absorption features of NH$_3$ are not observed) -- reaching spectrally-flat in the visible range. This makes the imaginary index of composition~\ref{comp:grey} intermediate between compositions~\ref{comp:ammonia} and~\ref{comp:saturn} for haze particles. 
 The refractive indexes for the four kind of haze particles are compared in Figure~\ref{fig:refractive_indexes}.
We note that NH$_4$SH particles (composition~\ref{comp:hydrosulf}) are expected to be brighter than the other kind of particles, as a result of the low real index of NH$_4$SH, while the ``Saturn'' particles (composition~\ref{comp:saturn}) should be the most absorbing ones in the visible, owing to their higher imaginary index shortward of 1~$\mu$m.

In order to study the impact of the cloud properties on the planetary albedo, we first perform 1-D radiative-convective simulations for globally-averaged conditions. The planetary albedo is defined as $1 - \frac{ASR}{ISR}$, where ISR stands for incoming solar radiation and ASR for absorbed solar radiation, both quantities being evaluated globally. This value of modeled planetary albedo is to be compared to the observed Bond albedo of 0.50 according to \citet{Li2018}. 
In this first set of simulations, the aerosol vertical structure is fixed with a reference cloud deck at 840~mbar and a scale height of 0.2 times the atmospheric scale height (to be consistent with previous observations of a compact cloud) along with an upper haze extending from 660 to 150~mbar, with a scale height set to the atmospheric scale height.
Table~\ref{tab1} presents a set of results for different couples of cloud and haze composition, varying the haze particle size between 0.5 and 2~$\mu$m, the haze integrated optical depth at 0.75~$\mu$m between 1 and 4, the cloud particle size between 10 and 20~$\mu$m, and the cloud integrated optical depth in the visible between 7 and 15 (only 22 out of the 108 combinations tested are shown in Table~\ref{tab1}).

Overall, we find that all cases considering a pure NH$_4$SH cloud lead to a too bright albedo ($>$0.55), regardless of the assumptions on cloud optical depth, cloud particle size or haze properties. Similarly, all cases with ``Saturn''-like haze particles combined with ammonia cloud particles result in too dark albedos ($\sim$0.4), which is consistent with the high imaginary index of these haze particles in the visible. A combination of a dark ``Saturn'' haze with a bright NH$_4$SH cloud leading to a $\sim$0.5 albedo might be found, but we choose to discard solutions with the ``Saturn'' haze as several studies \citep[e.g.,][]{Tomasko1978} suggest that Jupiter's haze particles must have a larger single scattering albedo than Saturn's. The ``grey'' and NH$_3$ haze particles considered here are in better agreement with estimates of the single scattering albedo by \citet{Tomasko1978}. 
The latter study also constrained the phase function of upper tropospheric cloud particles, which were found to have a strong forward scattering lobe in the visible.
The asymetry parameter (computed from a Mie scattering code) of our cloud particles lies in the range 0.8 to 0.85, which also indicate strong forward scattering.

Albedos comparable to that reported by \citet{Li2018} are obtained for combinations of ``grey'' and/or  NH$_3$  particles for the haze and cloud material, with the condition that the haze optical depth amounts to 3--4. Different combinations of the nature of the haze and cloud particles (NH$_3$--NH$_3$, grey--grey, grey--NH$_3$) give similar results, which is not surprising given their similar optical constants.  
Hence, even though spectroscopic studies have ruled out pure NH$_3$ ice particles for the cloud composition, it seems that assuming a NH$_3$ or "grey" cloud does not impact greatly the overall energy budget, and our model results are not very sensitive to one or the other type of composition.
We confirm that small haze particles ($\sim$0.5~$\mu$m) are needed in order to reproduce the 3 to 4 times larger haze optical depth observed in the visible compared to the near infrared: with the haze optical depth set to 4 at 0.75 $\mu$m for ammonia or grey particles, the optical depth amounts to $\sim$~1 at a wavelength of 2~$\mu$m, which is compatible with observations by \citet{Irwin2001}.

\begin{table*}
  \centering
  \caption{Cloud and haze properties along with the planetary albedo computed from globally-averaged 1-D radiative-convective simulations. Bold figures highlight our favored scenario, for which the albedo is close to 0.5, as reported by \citet{Li2018}. }
    \begin{tabular}{lcclccc}
    \toprule
    
    Cloud & & & Haze & & & \\
    \cmidrule{1-6}
	 type & size & $\tau$ cloud & type & size & $\tau$ haze &  Bond \\
   & (in $\mu$m)  & at 750 nm &   & (in $\mu$m)& at 750 nm   & Albedo \\
    \midrule

   NH$_4$SH & 15.00 & 7.00 & Grey  & 0.50 & 2.00 & 0.59\\ 
   NH$_4$SH & 15.00 & 7.00 & Grey  & 1.00 & 2.00 & 0.59\\ 
   NH$_4$SH & 15.00 & 7.00 &  Grey  & 0.50 & 4.00 & 0.61\\ 
   NH$_4$SH & 15.00 & 10.00 & Grey  & 0.50 & 2.00 & 0.63 \\ 
   NH$_4$SH & 15.00 & 7.00 & NH$_3$ & 1.00 & 2.00 & 0.59\\ 
   NH$_4$SH & 15.00 & 7.00 & ``Saturn''  & 1.00 & 2.00 & 0.55 \\
   NH$_4$SH & 15.00 & 15.00 & ``Saturn''  & 1.00 & 2.00 & 0.62 \\
   \midrule

   NH$_3$ & 15.00 & 7.00  & ``Saturn''  & 1.00 & 2.00 & 0.39 \\
   NH$_3$ & 15.00 & 7.00  & ``Saturn''  & 1.00 & 4.00 & 0.43 \\
   NH$_3$ & 15.00 & 15.00 & ``Saturn''  & 1.00 & 2.00 & 0.40 \\
   NH$_3$ & 15.00 & 15.00 & ``Saturn''  & 1.00 & 4.00 & 0.43 \\
   \midrule

   NH$_3$ & 15.00 & 4.00 & Grey  & 0.50 & 2.00 &  0.41 \\ 
   NH$_3$ & 15.00 & 10.00 & Grey  & 0.50 & 2.00 &  0.43 \\ 
   NH$_3$ & 15.00 & 7.00 & Grey  & 1.00 & 2.00 &  0.42 \\
   NH$_3$ & 15.00 & 7.00 & Grey  & 2.00 & 2.00 &  0.42 \\
   NH$_3$ & 10.00 & 7.00 & Grey  & 0.50 & 2.00 &  0.44 \\ 
   NH$_3$ & 20.00 & 7.00 & Grey  & 0.50 & 2.00 &  0.41 \\ 
      \midrule

   NH$_3$ & 15.00 & 7.00 & Grey  & 0.50 & 4.00 &  \textbf{0.48} \\ 
   NH$_3$ & 10.00 & 10.00 & Grey  & 0.50 & 4.00 &  \textbf{0.50} \\
   NH$_3$ & 10.00 & 15.00 & Grey  & 0.50 & 4.00 &  \textbf{0.51} \\
   NH$_3$ & 10.00 & 10.00 & NH$_3$ & 0.50 & 4.00 & \textbf{0.51} \\
   Grey   & 10.00 & 10.00 & Grey  & 0.50 & 4.00 &  \textbf{0.48} \\


    \bottomrule
    \end{tabular}
  \label{tab1}

\end{table*}

\begin{figure}
  \centering
  \includegraphics[width=0.8\linewidth]{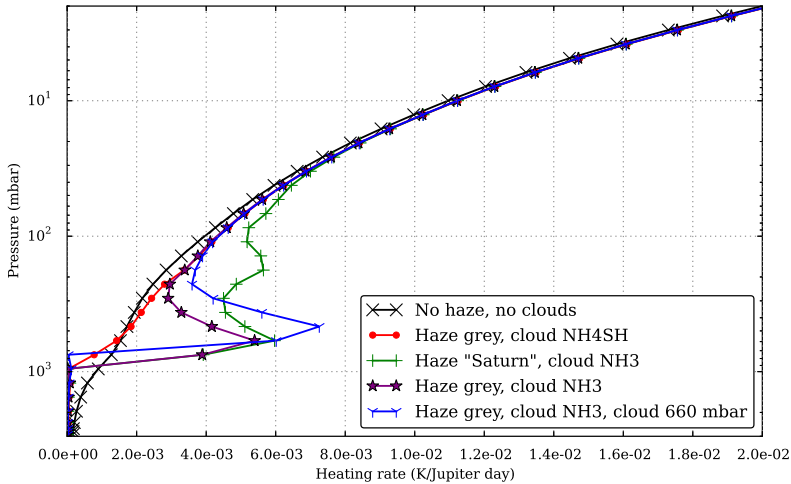}
  \caption{Vertical profiles of heating rates (in Kelvin per Jupiter day) due to absorption of solar radiation in the visible and near infrared for different cloud and haze particles, for globally-averaged conditions and Ls=180°. In this example, the haze and cloud optical depth at 0.75~$\mu$m are set to 4 and 10, respectively, and the haze and cloud particle sizes to 0.5 and 10~$\mu$m. The cloud deck is set at 840~mbar except for one case with a slightly shallower cloud deck, in blue.}
  \label{fig:heating_rate}
\end{figure}

The heat deposition differs depending on the cloud and haze composition, integrated optical depth and altitude of the cloud deck, as illustrated in Fig~\ref{fig:heating_rate}. For scenarios with bright NH$_4$SH clouds, the heating rate decreases moderately within the haze and cloud layer, while for scenarios with ammonia or ``grey'' cloud particles, the heat deposition reaches a local maximum within the cloud layer. Actually, because the cloud optical depth is large, the maximum heat deposition occurs above the cloud deck. In other words, at the cloud deck, there is little visible radiation left to be absorbed. Fig~\ref{fig:heating_rate} also illustrates the larger heating rate resulting from the absorption by ``Saturn''-like haze particles compared to ``grey'' particles. We also note that excluding completely haze and cloud opacity in the model results in unrealistic albedo (0.15) and heat deposition profile.

We then evaluate the impact of changing the cloud optical depth as well as the altitude of the cloud deck on the temperature. In our ``grey haze, NH$_3$ cloud'' scenario, increasing the cloud optical depth from 7 to 15 results in a 3~K warming of the troposphere (below the 300-mbar level), as absorption of visible solar photons increases. As stated in the introduction, we note that the opposite trend is actually observed on Jupiter: the cloudy equatorial zone is found to be 2 to 5~K colder than warmer, less cloudy, equatorial belts \citep[e.g.,][]{Fletcher2016}). This reinforces the idea that zones are regions of upwelling \citep[see for instance][]{Gierasch1986} where adiabatic cooling dominates over radiative heating.
Finally, moving the cloud deck from 840 to 660 mbar results in warming the troposphere by 5~K (while having little impact on the albedo, of the order of a 2\% change), as more solar light is absorbed at higher altitudes. 
Because the resulting temperature at the 1-bar level is closer to observations (detailed in section~\ref{sec:temperature}) when setting the cloud deck at 840~mbar, we choose to keep this setting throughout this paper.

In this section, we have documented the impact of different cloud and haze scenario on the tropospheric temperature and albedo. One has to keep in mind that modifications of the cloud or haze optical depth and altitude distribution will influence the tropospheric temperature by a few kelvins, just like modifications of the ammonia and hydrocarbon mixing ratio will also influence the temperature. Setting a realistic meridional profile of these variables is beyond the scope of the current project: not only observational constraints are limited, but feedbacks with meridional circulation -- not yet included -- are expected to play an important role as well. 
Hence, in the goal of setting an effective parametrization, able to reproduce the mean tropospheric temperature and global albedo, our nominal scenario is the following: a haze layer with an integrated optical depth of 4 in the range 660--150~mbar, ``grey'' particles of radius 0.5 $\mu$m on top of a NH$_3$ cloud (or indifferently a "grey" cloud) with 10-$\mu$m particles, a cloud deck at 840~mbar with an integrated optical depth at 750~nm of 15.

\subsection{Stratospheric aerosols}
\label{sec:aero_aurora}

\subsubsection{Observational constraints and motivation}
In addition to the tropospheric cloud and aerosol layers described above, we take into account two stratospheric haze layers: 
\begin{enumerate}
    \item We include an optically thin stratospheric layer comprising small spherical particles (0.2--0.5 $\mu$m) with an integrated optical depth set to 0.02 in the NIR and UV, as constrained by \citet{Zhang2013}. Their refractive index have been constrained in the same study, with imaginary parts at 255~nm and 900~nm estimated to 0.02 and 0.001, respectively. This haze is uniform with latitude and extends from the tropopause to the upper stratosphere, with a scale height equal to the atmospheric scale height. 
    Its impact on the stratospheric temperature is $<$0.5K.
    \item We include another layer that is not uniform with latitude and is more absorbant in particular in the UV, described further below.
\end{enumerate}
The addition of this second kind of aerosol is motivated by the observations of dark polar hoods at near-UV wavelength \citep{Hord1979, Tomasko1986}, which have been attributed to a stratospheric haze layer. This haze is both forward scattering and strongly polarizing, which implies that it is composed of aggregate particles similar to Titan's haze particles \citep{West1991}.
 The favored scenario for their formation is through precipitation of energetic particles in Jupiter's upper atmosphere in its auroral regions \citep{Pryor1991}, thought to be responsible for the production of heavy hydrocarbons \citep{Wong2003}. According to chemical and microphysical models, these hydrocarbons can condense and form fractal aggregates through coagulation processes \citep{Friedson2002}. 
 
 Recently, \citet{Zhang2013} brought new constraints on the size, shape, vertical and meridional distribution of these stratospheric aerosols by combining ground-based near-IR spectra from \citet{Banfield1998} and multiple phase angle images from the Cassini Imaging Science Subsysteme (ISS). The authors first derive the vertical profile of the aerosol mixing ratio at different latitudes and find that the stratospheric haze layer resides at a pressure of 50~mbar at low latitudes and $\sim$20~mbar at high latitudes (60--70\textdegree). Regarding the aerosol sizes and shapes, ISS observations can be fitted with small sub-micron spherical particles at low latitudes (which corresponds to the first type of stratospheric haze layer included in our model). Poleward of 30\textdegree N and 45\textdegree S,  ISS observations are consistent with fractal aggregates with a fractal dimension of 2, corresponding to an effective radius of about 0.7$\mu$m.
\citet{Zhang2013} also constrain the real and imaginary part of the refractive index of the fractal aggregates at two wavelengths in the UV and near-IR (at 255~nm and 900~nm).
They derive a family of solutions, with different plausible combinations of refractive indexes, number and radius of monomers.
For instance, their reference case corresponds to an imaginary index $n_i$ of 0.02 at 255~nm and 10$^{-3}$ at 900~nm and aggregates comprising 1000 monomers with a 10-nm radius. Other solutions  can match the Cassini/ISS observations. Two extreme cases are : aggregates comprising 100 monomers of 40~nm, with higher optical refractive indexes (0.08 in the UV and 5$\times$10$^{-3}$ in the NIR); or aggregates comprising 10000 monomers of 5~nm, with lower optical indexes (6.10$^{-3}$ in the UV and 2.10$^{-4}$ in the NIR).

Based on these observational constraints, \citet{Zhang2015} show that this haze dominates the radiative heating budget at middle and high latitudes in  Jupiter's middle stratosphere, with a contribution of the haze reaching up to 10 times the heating rate due to CH$_4$ alone in the 10--20~mbar pressure range. 
This haze can also cool the atmosphere through its infrared emission (see also \citet{Guerlet2015} for a  Saturn counterpart).
Hence, radiative heating and cooling by the polar haze appears to be a key component to be included in any radiative-convective equilibrium model of Jupiter. The parametrization of this haze in our model is detailed below.

\subsubsection{Parametrization of the aerosol properties}

\begin{figure}
  \centering
  \includegraphics[width=0.7\linewidth]{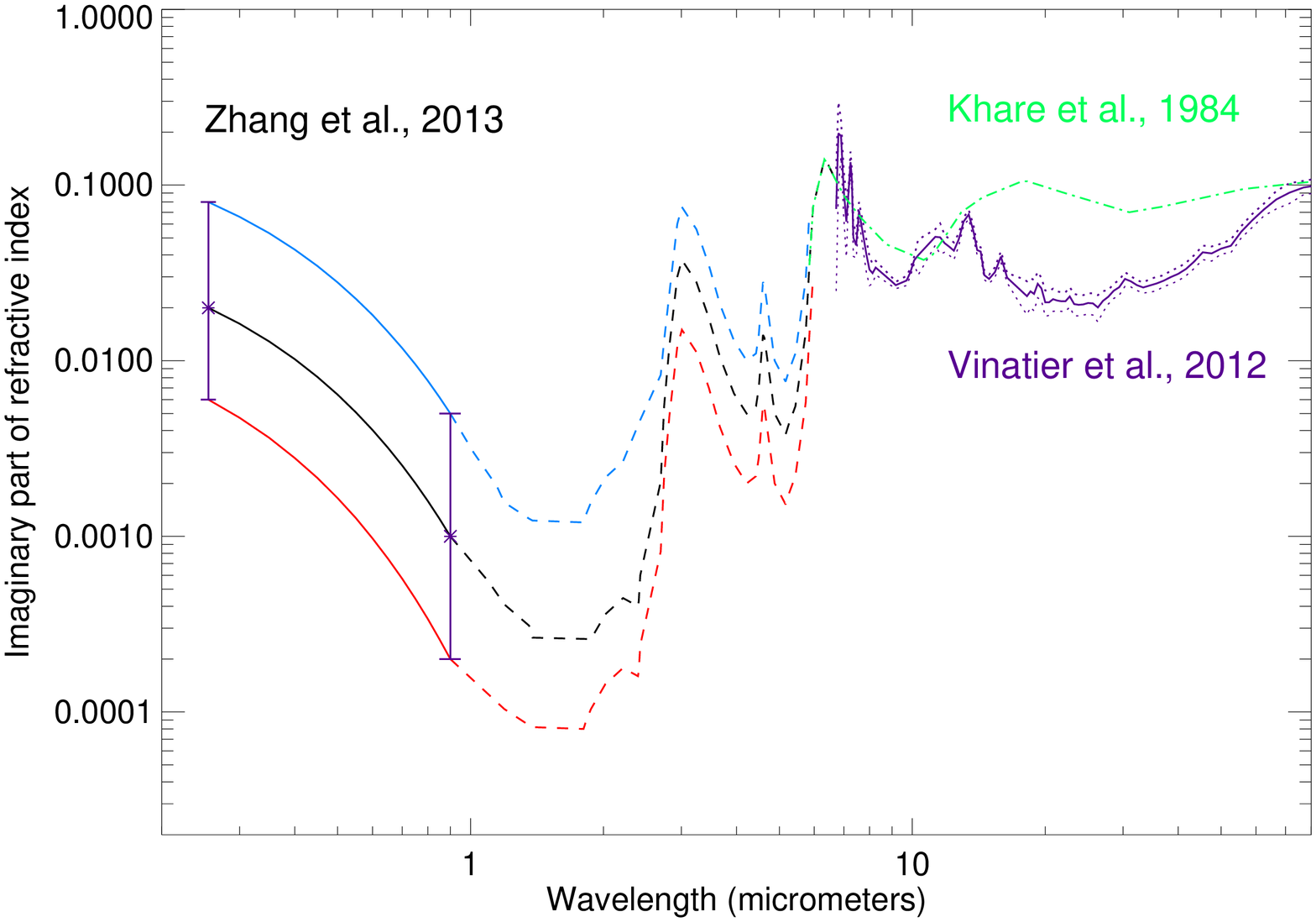}
  \caption{This figure illustrates the imaginary indexes used for generating optical properties of the stratospheric polar haze. Four sets of optical properties were computed based on different imaginary index, radius and number of monomers. We adopt the values constrained by \citet{Zhang2013} at NIR and UV wavelength, and \citet{Vinatier2012} (in purple) or \citet{Khare1984} (divided by 2 here, green line) in the thermal infrared. Between 1 and 7~$\mu$m, due to the lack of observing constraints, these values are interpolated, following the wavelength-dependence derived from tholins experiments by \citet{Khare1984}.}
  \label{fig:index_polar_haze}
\end{figure}

The optical properties  (extinction coefficient, scattering albedo and asymmetry factor) of fractal aggregates haze particles are computed using a semi-empirical model from \citet{Botet1997}. This model employs the mean-field approximation  in the case of scattering of an electromagnetic wave by a cluster of monosized spheres. 
We compute the optical properties for aggregates with a fractal dimension of 2, and for the three aforementioned scenario determined by \citet{Zhang2013} regarding the number and radius of monomers. 
From UV to NIR, the real index is set to 1.65, similar to the mean value of \citet{Zhang2015}. The imaginary refractive indexes were set to the three set of values determined by \citet{Zhang2013} at 255 and 900~nm -- one for each combination of number and radius of monomers --  with a logarithmic interpolation in between as in \citet{Zhang2015}.
In the thermal infrared, due to the lack of observational constraints, we adopt the real and imaginary index of \citet{Vinatier2012} derived from Cassini/CIRS observations of Titan's stratospheric haze, which present striking similarities with Saturn's polar haze \citep{Guerlet2015}. For this given set of refractive index, optical properties in the thermal infrared are computed three times, for the three aforementioned values of number and radius of monomers. This ensures consistency in the considered scenario.
Finally, we also include a case where tholins-like properties are assumed \citep{Khare1984}. They are not considered a very relevant analog for Jupiter's haze -- they do not match the observed properties of Titan's haze \citep{Vinatier2012}, let alone Saturn's haze -- but this test is useful for sensitivity studies. For this test, we only consider the scenario with 1000 monomers of 10-nm radius and we actually divide by 2 the refractive index derived from the laboratory experiments of \citet{Khare1984} to better match observations by \citet{Vinatier2012}. 
These different refractive index are summarized in figure~\ref{fig:index_polar_haze} and our four scenarios summarized in table~\ref{table:strato_haze}.
We will discuss the impact of these different properties on the thermal structure in the next section.

\begin{table*}
  \centering
  \caption{Description of the 4 sets of parameters used to generate the stratospheric polar haze optical properties. N$_{mono}$ is the number of monomers in the aggregates, r$_{mono}$ the radius of the monomers, Im stands for the imaginary index.}
    \begin{tabular}{llllll}
    \hline
	 Scenario & N$_{mono}$ & r$_{mono}$ &  Im(255~nm) & Im(900~nm) & Im(Thermal infrared)   \\
    \hline

   1 & 10000 & 5~nm & 6.10$^{-3}$  & 2.10$^{-4}$  & \citet{Vinatier2012}\\ 
   2 & 1000 & 10~nm & 2.10$^{-2}$  & 1.10$^{-3}$  & \citet{Vinatier2012}\\ 
   3 & 100 & 40~nm & 8.10$^{-2}$  & 5.10$^{-3}$  & \citet{Vinatier2012}\\ 
   4 & 1000 & 10~nm & 2.10$^{-2}$  & 1.10$^{-3}$  & \citet{Khare1984}\\ 
 
    \hline
    \end{tabular}
  \label{table:strato_haze}
\end{table*}

\begin{figure}
  \centering
  \includegraphics[width=0.48\linewidth]{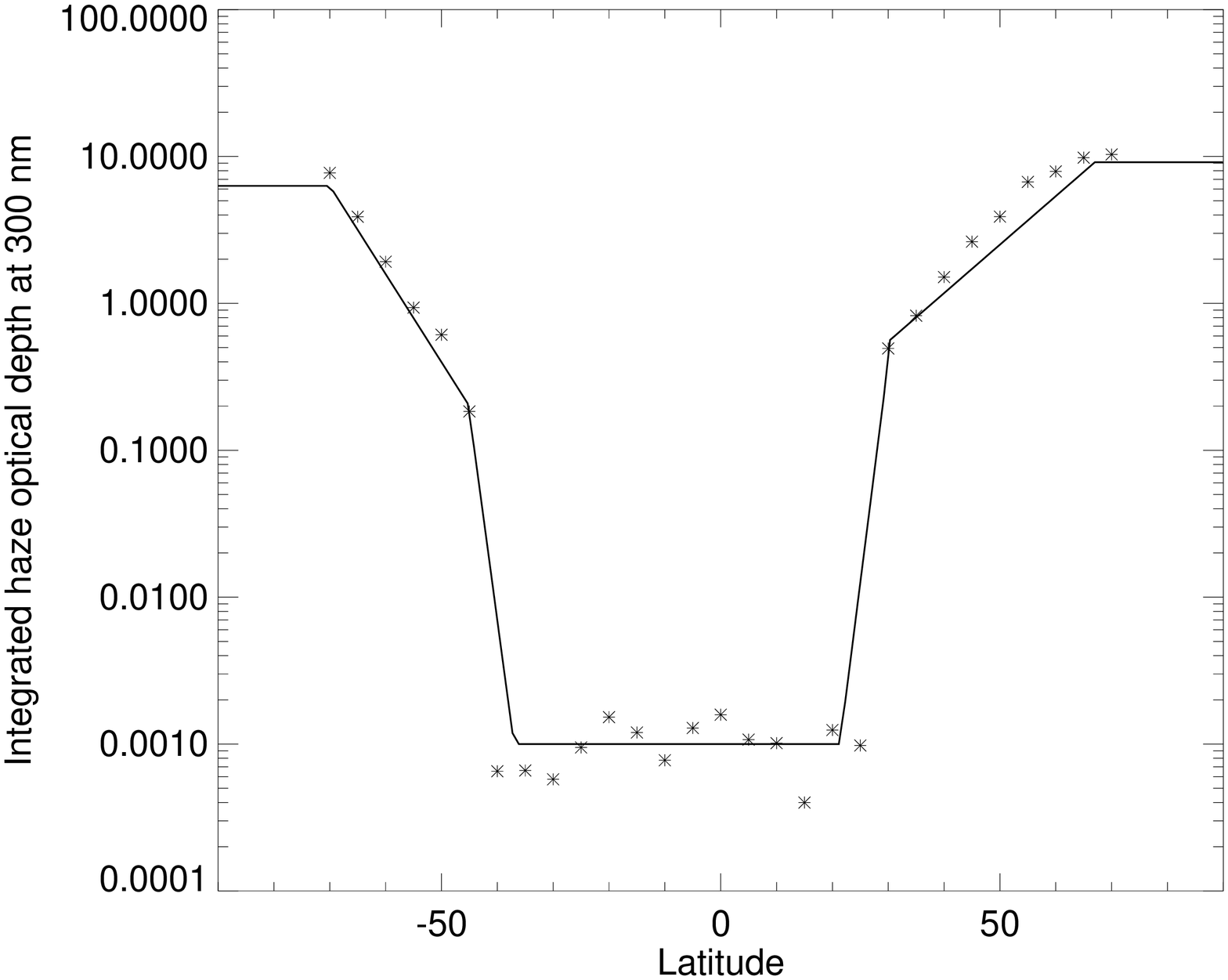}
  \includegraphics[width=0.48\linewidth]{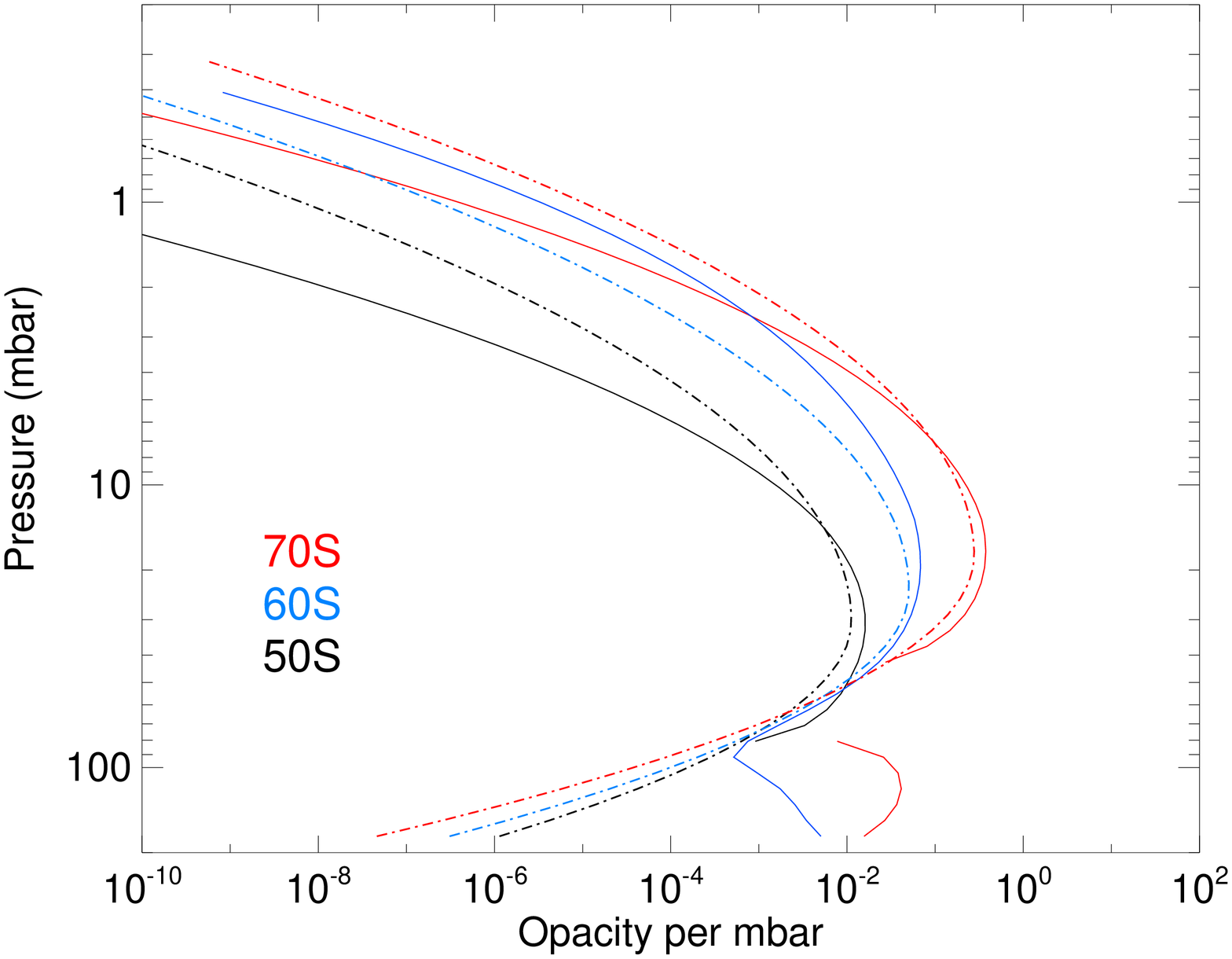}
  \caption{Left: Meridional variation of the stratospheric haze optical thickness at 300~nm integrated above the 80~mbar level as derived from \citet{Zhang2013} (stars) compared to the values  adopted in our model (dashed line). Right: Opacity vertical profiles at 300~nm derived from \citet{Zhang2013} retrieved number density profiles, shown at three latitudes (solid lines), compared with those adopted in our model (dashed lines).}
  \label{fig:aero_zhang}
\end{figure}

Regarding the meridional variations of the polar haze optical thickness, we build a meridional profile based on that retrieved by \citet{Zhang2013}, with a slightly smoother transition  at mid-latitudes where the optical depth varies by several orders of magnitude over a few degrees of latitude. The integrated haze optical thickness as a function of latitude adopted in our model is compared with \citet{Zhang2013} retrievals in Figure~\ref{fig:aero_zhang}. Poleward of 70\textdegree\  \citep[the highest latitude observed in ][]{Zhang2013}, we simply assume that the haze optical depth is equal to that at 70\textdegree.

We choose to parameterize the aerosol opacity vertical profile  with a skewed gaussian profile:
\begin{equation}
\label{eq:profile_aurora}
\tau(p) \propto exp(-(H \times ln(p/p_m))^2 )/2 \delta_1^2)/(\delta_1+\delta_2)
\end{equation}
with $\tau$ the opacity per mbar, $p$ the pressure, $p_m$ parametrizing the pressure level where the optical depth is maximum and $\delta_1$ and $\delta_2$ parametrizing the width and skewness of the profile.
This function reproduces well, to first order, the vertical profile of the haze opacity derived from number density profiles retrieved by \citet{Zhang2013} (see  Fig.~\ref{fig:aero_zhang}). We vary $p_m$ linearly with latitude to capture the fact that the haze layer shifts from $\sim$40~mbar to 20~mbar between $\sim$45 and 70\textdegree, as derived  by \citet{Zhang2013}. In our model, the opacity profile is then normalized at each latitude bin so that the optical depth integrated above the 80~mbar level matches the meridional profile in Fig.~\ref{fig:aero_zhang}. 
 The resulting vertical profiles of the haze opacity are shown  at three latitudes in Figure~\ref{fig:aero_zhang}, along with those retrieved by \citet{Zhang2013}. 
 We note that \citet{Zhang2013} retrievals suggest that, poleward of 60\textdegree, the tropospheric aerosol layer shifts to lower pressure levels ($\sim$100 mbar instead of 200~mbar), which we did not take into account (our tropospheric layer extends up to 180 mbar at all latitudes).

\subsubsection{Radiative impact of the haze}

\begin{figure}
  \centering
  \includegraphics[width=0.48\linewidth]{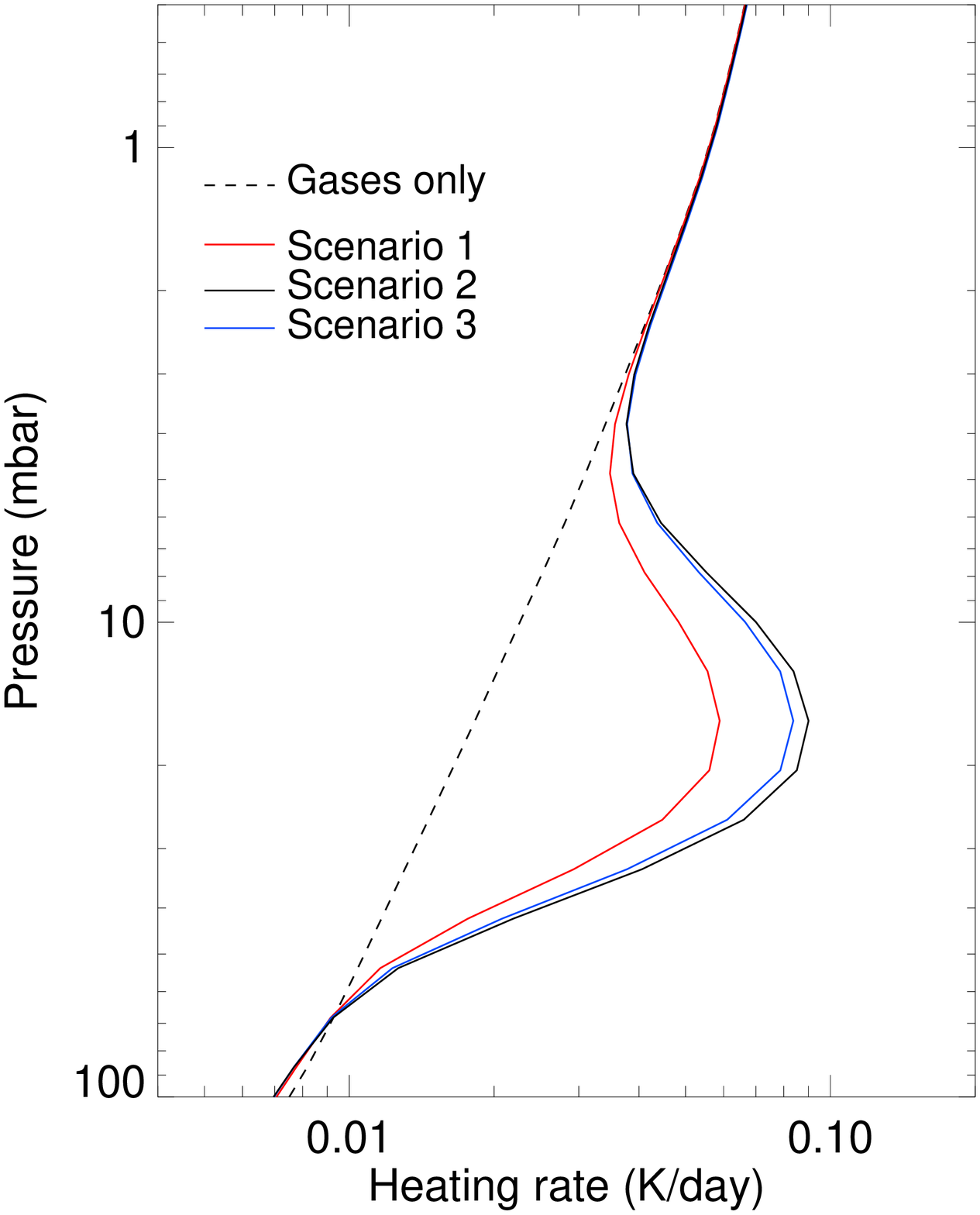}
  \includegraphics[width=0.48\linewidth]{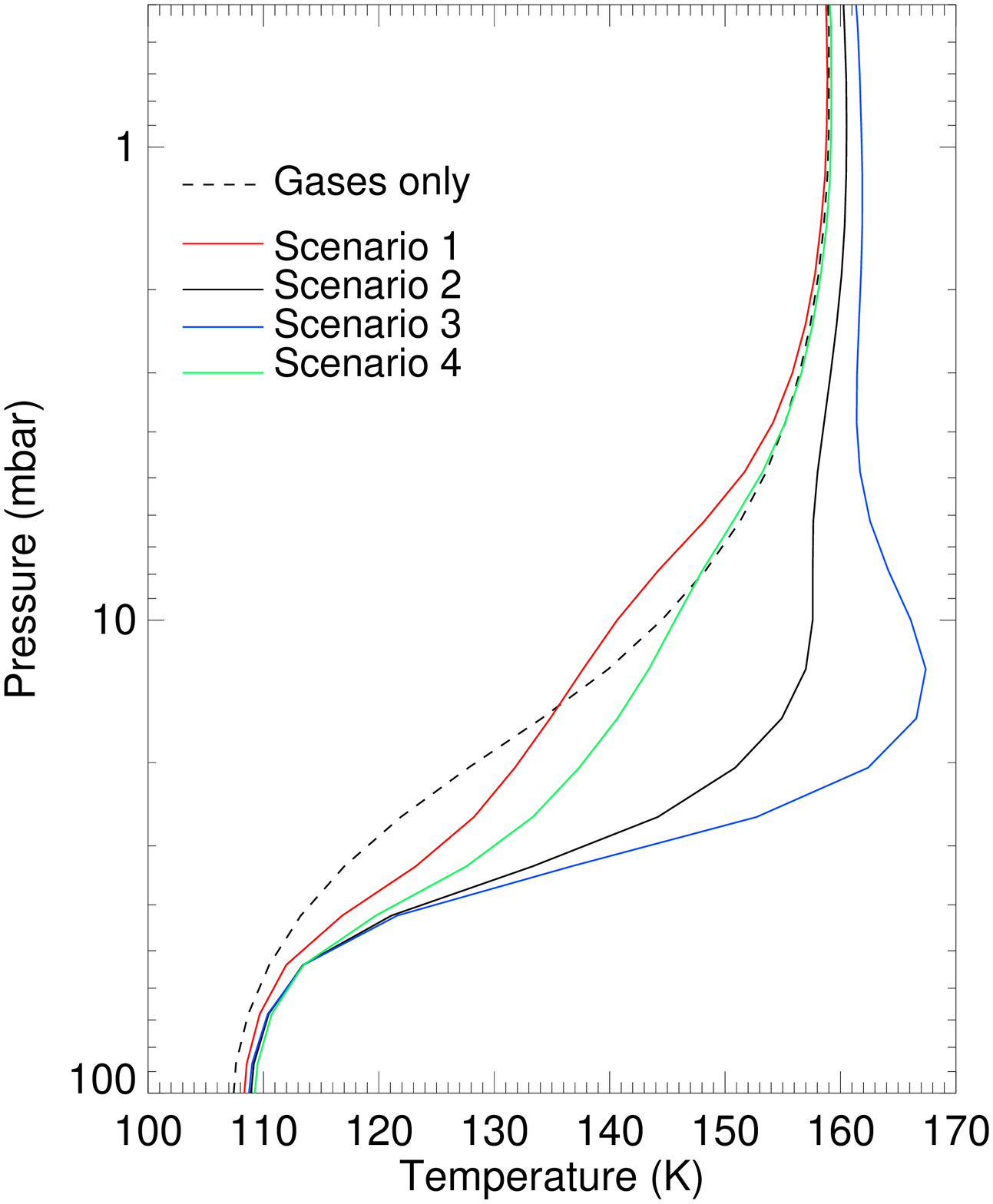}
  \caption{Daily-averaged heating rate, in Kelvin per Earth day (left) and temperature (right) vertical profiles at latitude 60\textdegree S and Ls= 0\textdegree . Dashed black lines correspond to a case without the polar haze contribution, colored lines refer to different polar haze scenarios described in table~\ref{table:strato_haze}. Scenario 4 is not shown in the heating rate figure as it is similar to scenario 2.}
  \label{fig:impact_aero_zhang}
\end{figure}

As reported by \citet{Zhang2015}, we confirm that including the polar haze results in strongly enhanced heating rates in the middle stratosphere, mostly between 40 and 70\textdegree N and 50 and 70\textdegree S. Figure~\ref{fig:impact_aero_zhang} shows an example at 60\textdegree S, Ls=  0\textdegree , where the heating rate is increased by a factor of four to six at the 10-mbar pressure level when stratospheric aerosols are included, which is in overall agreement with the 5 to 10 times enhancement factor reported by \citet{Zhang2015}. Using our radiative-convective equilibrium model, we can go further and evaluate for the first time the impact of the polar haze on the equilibrium temperature profile. 
We find that the temperature is very sensitive to the polar haze properties. At this latitude and season, considering scenario 2 or 3, the impact of the polar haze is to significantly warm the stratosphere by 20K to 30K in the 10 to 30-mbar pressure range (see Figure~\ref{fig:impact_aero_zhang}). This effect decreases with altitude, amounting to 3--5K at the 1-mbar pressure level.
If scenario 4 is considered (same as scenario 2 but with the imaginary index of \citet{Khare1984} divided by 2, more absorbant in the thermal infrared than that of \citet{Vinatier2012}), the polar haze net effect is a moderate warming (10K) that is maximum near the 30-mbar pressure level. 
Finally, scenario 1 results in temperatures changes of the order of +5K near the 30-mbar level and -5K near the 5-mbar level. Indeed, in this case, a net cooling of the middle stratosphere is obtained despite the increase in heating rates. This is explained by the simultaneous increase in cooling rates due to the polar haze.
We also note that despite similar heating rates in scenario 2 and 3, the equilibrium temperature profiles differ significantly. There again, these differences result from different cooling rates: even though the thermal infrared imaginary indexes of \citet{Vinatier2012} are used in both scenario 2 and 3, the number and radius of monomers was varied among these scenario.
A smaller number of larger monomers (scenario 3) is more efficient in cooling the atmosphere than a larger number of smaller monomers (scenario 2).
Hence, it appears crucial to better characterize the haze properties (their refractive index but also size and dimensions of the aggregates) in order to realistically model their radiative impact.

At high latitudes ($>$ 75\textdegree) the net radiative impact of the polar haze depends on season: during winter, the net effect of scenario 2 is to cool the atmosphere, by typically 10~K between 20 and 2~mbar. This is easily explained by the fact that the solar insolation is near zero at this season and location, while the aerosol layer still emits longwave radiation. On the other end, over the summer pole, the net effect can be an important warming (10--15K)  between 20 and 2~mbar. Comparisons with observations, using simulations including or not the polar haze, are shown in the next section.

\section{Thermal structure and comparisons to observations}
\label{sec:temperature}

\subsection{Internal heat flux and tropospheric temperature}
\label{sec:heatflux}

Before presenting in detail the results of our radiative-convective model at equilibrium, we address the issue of the tropospheric equator-to-pole temperature gradient. Indeed, it is well  known since the Pioneer 10 and 11 era that Jupiter exhibits no significant latitudinal gradient of temperature or emitted thermal infrared flux   at the 1-bar level \citep{Ingersoll1976}.
However, assuming a uniform internal heat flux $F_{cst}=7.48$~W.m$^{-2}$, our radiative-convective model produces a strong temperature contrast of 28~K at 1 bar between the warmer equator (178~K) and colder poles (150~K) (see Figure~\ref{fig:compa_fletcher_tropo}). This is expected from such a radiative model, as solar insolation is maximum at low latitudes all year round (given Jupiter's low obliquity). 

To explain the observed near-uniform tropospheric temperatures, several theories have been proposed. For instance, using a turbulent, 3-D deep convection model, \citet{Aurnou2008} finds that convective heat transfer by quasi-geostrophic thermal plumes results in an outward heat flow 2.5 to 3.2 times greater at the poles than at equator. This latitudinal trend is consistent with the work of \citet{Pirraglia1984} who tried to estimate the meridional variations of internal heat flux needed to reconcile the observed solar energy deposition with the outgoing thermal radiation.
On a different note, with their General Circulation Model for Jupiter's troposphere, \citet{Young2019} found that even when considering a uniform internal heat flux, atmospheric dynamics acts to balance the latitudinal-varying solar forcing. As a consequence, the 1-bar equator-to-pole temperature gradient is reduced from 35~K with a radiative-convective version of their GCM to only 5~K when using their full GCM with resolved atmospheric dynamics.

In order to emulate these effects in our radiative-convective model, we test different functions to vary the internal heat flux $F_{int}$ with latitude $\theta$, for instance:
\begin{equation}
\label{eq:Fint1}
    F_{int} (\theta) = 0.67\times F_{cst} + 0.66 \times F_{cst} \times \sin^2(\theta)
\end{equation}
\begin{equation}
\label{eq:Fint2}
    F_{int} (\theta) = 0.5\times F_{cst} + F_{cst} \times \sin^2 \theta
\end{equation}
This \textit{ad hoc} parametrization ensures a planet-average internal heat flux equal to $F_{cst}$ while setting an internal heat flux twice larger (eq.~\ref{eq:Fint1}) or three times larger (eq.~\ref{eq:Fint2}) at the poles than at the equator.
When using eq.~\ref{eq:Fint1}, the equatorial temperature at the 1-bar level is now 9~K warmer than the poles (instead of 28~K when a uniform internal heat flux is assumed). The associated outgoing thermal emission is only 8\% larger at the equator than at 60\textdegree \ latitude, which is consistent with \citet{Pirraglia1984} observations, which extended to 60\textdegree \ only. 
However, when using eq.~\ref{eq:Fint2}, the temperature is actually 2~K warmer (and the outgoing thermal emission 8\% larger) at 60\textdegree \ than at the equator.
Hence, in what follows, we discuss the thermal structure obtained with eq.~\ref{eq:Fint1} which yields more realistic results. It is worth emphasizing here that the temperature field at pressures lower than $\sim$50~mbar is not impacted by the hypothesis of a uniform or varying internal heat flux.





\subsection{Thermal structure and seasonal trends}
\label{sec:seasons}

\begin{figure}
  \centering
  \includegraphics[width=0.95\linewidth]{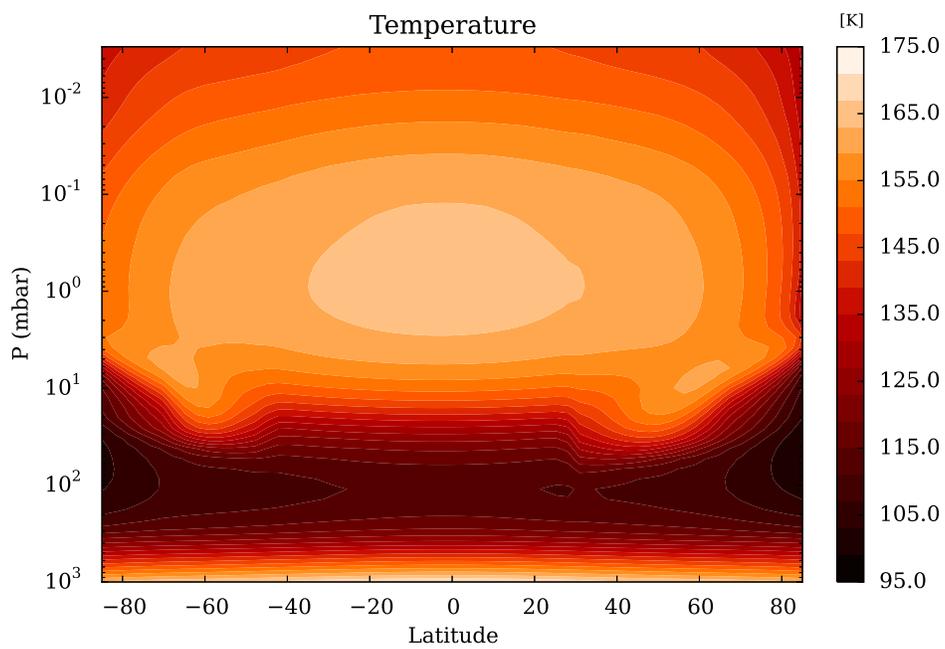}
  \caption{Pressure-latitude cross-section of the temperature (in K) in Jupiter's atmosphere at Ls=0\textdegree . The internal heat flux varies with latitude as defined in eq.~\ref{eq:Fint1} and the polar haze scenario~\#2 was used.}
  \label{fig:tempe_map_Ls0}
\end{figure}

We run our seasonal radiative-convective 1-D model on 32 distinct columns, each for a different latitude, and for 10 Jupiter years in order to reach radiative-convective equilibrium.
We performed three runs corresponding to polar haze scenario~\#1, 2 and 3.
In this section, we mostly present and discuss the results obtained with scenario~\#2, as we will see in section~\ref{sec:compa_obs} that it appears more consistent with observations.
The corresponding latitude-pressure cross-section of the temperature obtained at Ls=0\textdegree ~with scenario~\#2 is shown in Figure~\ref{fig:tempe_map_Ls0}. 
 From low- to mid-latitudes, our model reproduces well the tropopause altitude (100~mbar) and temperature (110~K) reported in previous studies \citep[e.g.][]{Conrath1998,Fletcher2016}. The stratospheric temperature is nearly isothermal in the range 3--0.1~mbar, where it reaches a maximum of 165~K. Above this level, the temperature decreases with altitude as infrared cooling dominates over solar heating. This is in agreement
with \citet{Kuroda2014} who find a 5~K temperature decrease between 0.1 and 0.001~mbar, from 160 to 155~K (ie. overall 5~K colder than our model predictions).
At latitudes 50\textdegree --70\textdegree , stratospheric temperatures are found to be colder than at low latitudes, except in the range 3--30~mbar where the warmer temperatures are due to the absorption of solar light by aerosols.
Equilibrium temperatures in this pressure and latitudinal range are strongly influenced by the assumed polar haze properties, as reported in section~\ref{sec:aero_aurora}. However, qualitatively, the thermal structure is similar regardless of the polar haze scenario. 
At latitudes poleward of 70\textdegree , temperatures are the coldest with a 100~K tropopause  and a maximum stratospheric temperature of 140~K to 150~K at the 1-mbar level. We note that at high latitudes, the tropopause is broader and extends from 100~mbar to 20~mbar, which is caused by the heating by CH$_4$ being less efficient in the lower stratosphere due to the low solar elevation.

\begin{figure}
  \centering
  \includegraphics[width=0.95\linewidth]{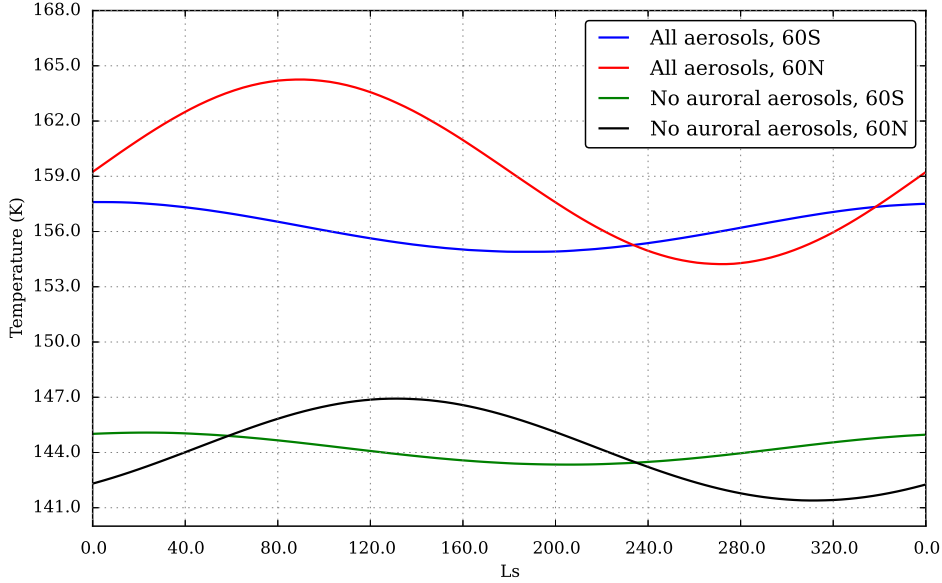}
  \caption{Temperature at the 10~mbar pressure level as a function of solar longitude (Ls, with Ls=0 corresponding to spring equinox in the northern hemisphere) for latitudes 60\textdegree N and 60\textdegree S, as labeled. Two cases are shown, including or not the stratospheric polar haze (scenario~\#2).}
  \label{fig:tempe_seasons}
\end{figure}

Seasonal variations are expected to be small owing to Jupiter's low obliquity. We present in Figure~\ref{fig:tempe_seasons} the seasonal evolution of the 10-mbar temperature at 60\textdegree N and 60\textdegree S, with and without polar haze (scenario~\#2). We first note that the amplitude of seasonal variations is very small in the southern hemisphere: it is only 2~K at 60\textdegree S, increasing to 3~K when the radiative impact of the polar haze is taken into account. 
This can be explained by the competing effects of obliquity and eccentricity, as Jupiter's perihelion occurs at Ls=57\textdegree \ close to southern "winter". On the other hand, these two effects add up in the northern hemisphere, where the peak-to-peak seasonal amplitude is $\sim$~6~K for the polar-haze-free case.
When polar aerosols are included  (scenario~\#2), there is a global temperature increase of 15~K at 60\textdegree N, 12~K at 60\textdegree S. The peak-to-peak amplitude of seasonal variations is also enhanced at 60\textdegree N  (10~K instead of 6~K) when we include this additional aerosol radiative forcing.
The seasonal amplitude reported here for 60\textdegree N and 10~mbar is similar should other pressure levels in the range  30~mbar and 0.01~mbar, and latitudes in the range  45\textdegree N -- 75\textdegree N, be considered.
Finally, at 60\textdegree N, when the polar haze are added, we also notice that the temperature maximum is shifted to an earlier season (Ls=95\textdegree~instead of Ls=125\textdegree), closer to northern summer solstice, which hints at shorter radiative timescales as a result of adding polar hazes (see next section~\ref{sec:rad_times} for further details).

\subsection{Radiative timescales}
\label{sec:rad_times}

In this section, we evaluate and discuss radiative relaxation timescales in Jupiter's atmosphere.  Radiative timescales can be used to assess whether the atmosphere responds quickly or not to changes in atmospheric temperatures and solar insolation. It is sometimes used in idealized global circulation models where radiative processes are parametrized with a relaxation scheme. Quantitative estimates of the characteristic radiative timescale of the jovian atmosphere have been rather limited in the past, as it requires a detail inventory of the radiative forcings, as is done in this study (see section~\ref{sec:model}). Recent estimates by \citet{Zhang2013b}, \citet{Kuroda2014} and \citet{ChengLi2018}, based on their respective radiative models, take into account gaseous radiative forcings similar to ours, but neglect any kind of clouds and aerosols. 

To compute the radiative relaxation timescales of Jupiter's atmosphere with our seasonal radiative-convective model, we adopt the following standard approach (see e.g. Eq.~6 in \citet{Kuroda2014}):  we run a 1-D radiative-convective simulation until radiative equilibrium is reached; then, we add 4~K to the resulting temperature profile at all levels and restart a simulation with this modified profile. Radiative relaxation time, $\tau_{rad}$, is obtained by dividing the temperature disturbance (here 4~K) by the change in net (daily-averaged) heating rates due to this disturbance. 

\begin{figure}
  \centering
  \includegraphics[width=0.55\linewidth]{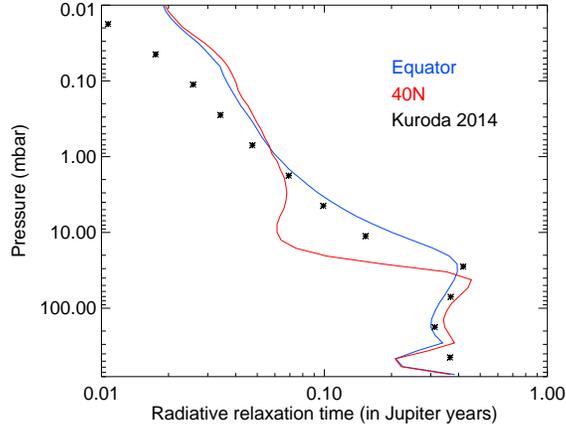}
  \caption{Two example profiles of the radiative timescale in Jupiter's atmosphere at L$_s$=0 at the equator (in blue) or at 40\textdegree N where polar hazes are abundant (in red, with scenario~\#2). These are compared to values published in \citet{Kuroda2014} for the equator (stars)}.
  \label{fig:rad_time}
\end{figure}

Two examples are shown in Figure~\ref{fig:rad_time} for latitudes 40\textdegree N and the equator. We find that in the upper troposphere, radiative timescales are of the order of 0.2 to 0.4 Jupiter years, meaning that any temperature disturbance due to, for instance, dynamical activity, can persist a long time before being equilibrated by radiative processes. 
In the stratosphere, this timescale shortens with altitude and is of the order of 3\% of a Jupiter year ($\sim$ 100 Earth days) at the 0.1~mbar level, meaning that a temperature disturbance will be radiatively damped over this timescale (if the source of the disturbance is not active anymore).
At the equator, we note that our radiative timescales  are in agreement with that derived by \citet{Kuroda2014}. Two notable exceptions are the upper stratosphere, where our timescales are about 50\% longer than in \citet{Kuroda2014}, and the lower troposphere, where our estimated timescale is twice shorter at the 500~mbar level. The former can be explained by the choice of slightly different hydrocarbon profiles at high altitudes and/or differences in spectroscopic calculations, and the latter by the lack of tropospheric aerosols in the model of \citet{Kuroda2014}. 
At 40\textdegree N and in the range 5--30~mbar, we find that the radiative timescale is two to five times shorter than at the equator. This is due to polar haze radiative forcing (here with scenario~\#2) and is consistent with our remark on seasonal temperature variations in section~\ref{sec:seasons}: at the 10-mbar level, the maximum of temperature occurs shortly after summer's solstice due to a quick response of the atmosphere to changes in solar insolation.
This feature was not captured by \citet{Kuroda2014} who neglected the radiative contributions of aerosols in their model.
All the conclusions in this paragraph hold when we compare our results to the similar work by \citet{Zhang2013b} and \citet{ChengLi2018}.

\subsection{Comparison to observations}
\label{sec:compa_obs}

The monitoring of Jupiter's spatio-temporal temperature variations from the analysis of thermal infrared spectra started with the Voyager spacecrafts in 1979 \citep[e.g.,][]{Hanel1979, Simon2006} and was later on followed by the Cassini fly-by of Jupiter in December, 2000 \citep[e.g.,][]{Flasar2004,Nixon2007}.
Jupiter's thermal structure has also been monitored very regularly from Earth-based telescopes, most notably from NASA's Infrared Telescope Facility (IRTF) \citep[e.g.,][]{Orton1991}. Nowadays, these studies are pursued using the Texas Echelon Cross Echelle Spectrograph (TEXES) instrument on the IRTF \citep{Lacy2002, Fletcher2016, Sinclair2017, Melin2018}.
This high-spectral resolution instrument is able to constrain the 3D temperature field in Jupiter's upper troposphere and stratosphere with a spatial resolution of 2--4\textdegree \ in latitude \citep{Fletcher2016}, which is actually comparable to the spatial resolution achieved by the Cassini fly-by. Both CIRS and TEXES are sensitive to the temperature in the pressure range 700--0.5~mbar (with the caveat of a low sensitivity in the 20--60~mbar range) and cover the latitude range 78\textdegree S -- 78\textdegree N.

In this section, we focus on the comparison with the results of \citet{Fletcher2016} who analysed spectra acquired by TEXES in December, 2014 (corresponding to Ls=175\textdegree ) and also analysed, with the same retrieval pipeline, observations from the Composite Infrared Spectrometer (CIRS) on board Cassini during the December, 2000 flyby  (corresponding to Ls=110\textdegree ).
In our comparisons, we neglect longitudinal variability and only consider temperatures derived from zonally-averaged spectra provided by \citet{Fletcher2016}.

\begin{figure}
  \centering
  \includegraphics[width=0.48\linewidth]{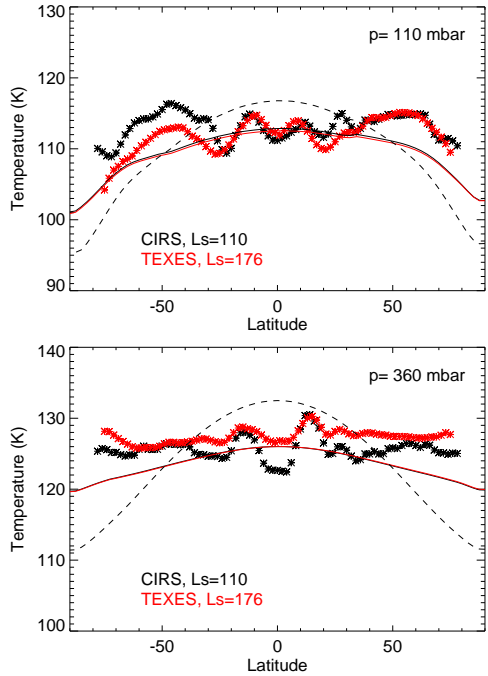}
 \caption{Comparison between tropospheric temperatures derived by \citet{Fletcher2016} from Cassini/CIRS (black stars) and TEXES (red stars) observations at two different seasons, as labeled, and that predicted by our model, in solid lines (in black for L$_s$=110, in red for  L$_s$=176). The upper and lower panels display temperatures at 110 and 360~mbar, respectively. These results are obtained with a latitudinal-varying internal heat flux; for reference, we also show the simulated temperature obtained when setting a constant internal heat flux (dashed line, L$_s$=110).}
  \label{fig:compa_fletcher_tropo}
\end{figure}

We first focus on the comparison in the upper troposphere, shown in Figure~\ref{fig:compa_fletcher_tropo}. At the 360-mbar level, the temperature derived from CIRS and TEXES shows little meridional or temporal variability except at the equator, where the TEXES-derived temperature is about 5~K warmer in 2014 than the CIRS-derived temperature in 2000. These variations are attributed to changes in the dynamics of the equatorial belts \citep{Fletcher2016}. As already mentioned in section~\ref{sec:cloud_sensi}, the cloudy equatorial zone is  colder than the less cloudy, warmer equatorial belts at 15\textdegree N and 15\textdegree S, which is thought to be the consequence of vertical motions (upwelling in zones, subsidence in belts) rather than due to a radiative effect.
Near the tropopause level (at 110~mbar), observed temperatures exhibit a small (5~K) decrease in  temperature from 50 to 78\textdegree \ in both hemispheres, and a temporal variability of the order of 3~K in the southern hemisphere. 

Our predicted temperatures, obtained with the parametrization of a latitudinal-varying internal heat flux (as is defined in eq.~\ref{eq:Fint1}) and a single cloud and haze scenario, reproduces reasonably well the observed globally-averaged temperature in the troposphere. Our main disagreement is that at the 360-mbar level, our temperatures are $\sim$4~K cooler than observations in the 40--78\textdegree \ latitude range. 
The fact that our model  still slightly underestimates the temperature at high latitudes suggests that we should let the internal heat flux increase even more with latitude. However, because this is only a crude parametrization that might become obsolete when full GCM simulations are run, we did not attempt to optimize the parameters of  eq.~\ref{eq:Fint1} until we obtained a perfect match with observations.
In addition, other processes could be at play in this model-observation mismatch: indeed, the aerosol opacity cross-section derived by \citet{Zhang2013} indicate that the tropospheric haze layer could extend at higher altitudes in the 60--70\textdegree \ latitude range (with no observations beyond 70\textdegree ), which could enhance the radiative heating in the upper troposphere. Given the current lack of observational constraints on the haze properties at high latitudes, we did not modify our tropospheric haze scenario.

\begin{figure}
  \centering
  \includegraphics[width=0.95\linewidth]{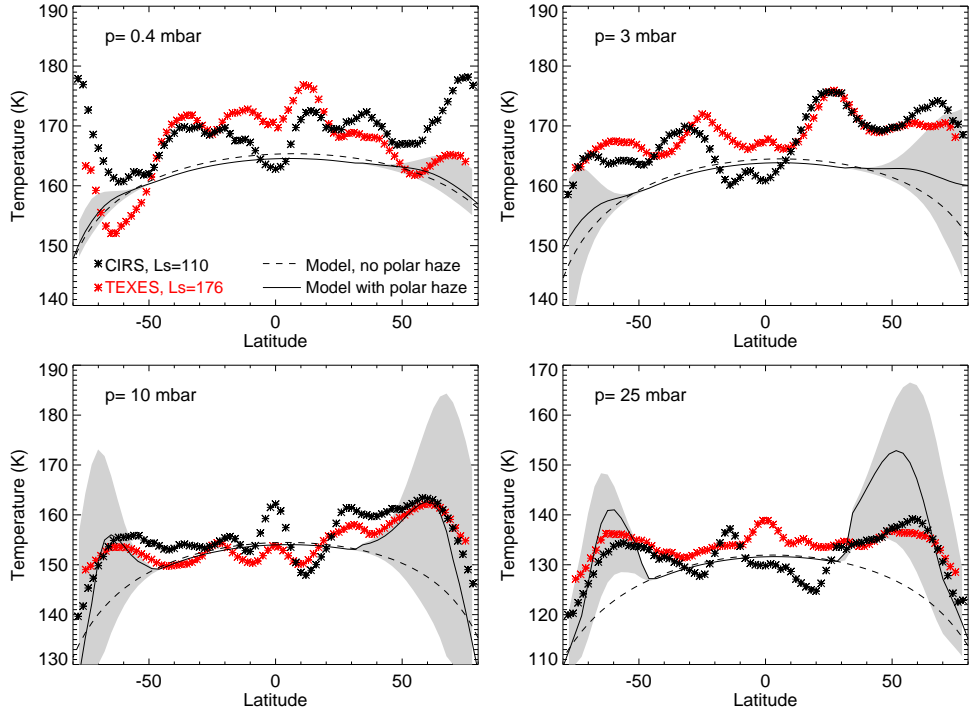}
  \caption{Comparison between stratospheric temperatures derived by \citet{Fletcher2016} from Cassini/CIRS and TEXES observations at two different seasons, as labeled (stars), and that predicted by our model at an intermediate season (Ls=140, lines). 
  The dashed line is for a case where the polar haze was neglected while the grey shading represents the effect of including polar haze scenarios~\#1 to 3 (with the solid black line referring to scenario~\#2). 
  The four panels display temperatures at four different pressure levels (0.4, 3, 10 and 25~mbar).}
  \label{fig:compa_fletcher_strato}
\end{figure}

\begin{figure}
  \centering
  \includegraphics[width=0.6\linewidth]{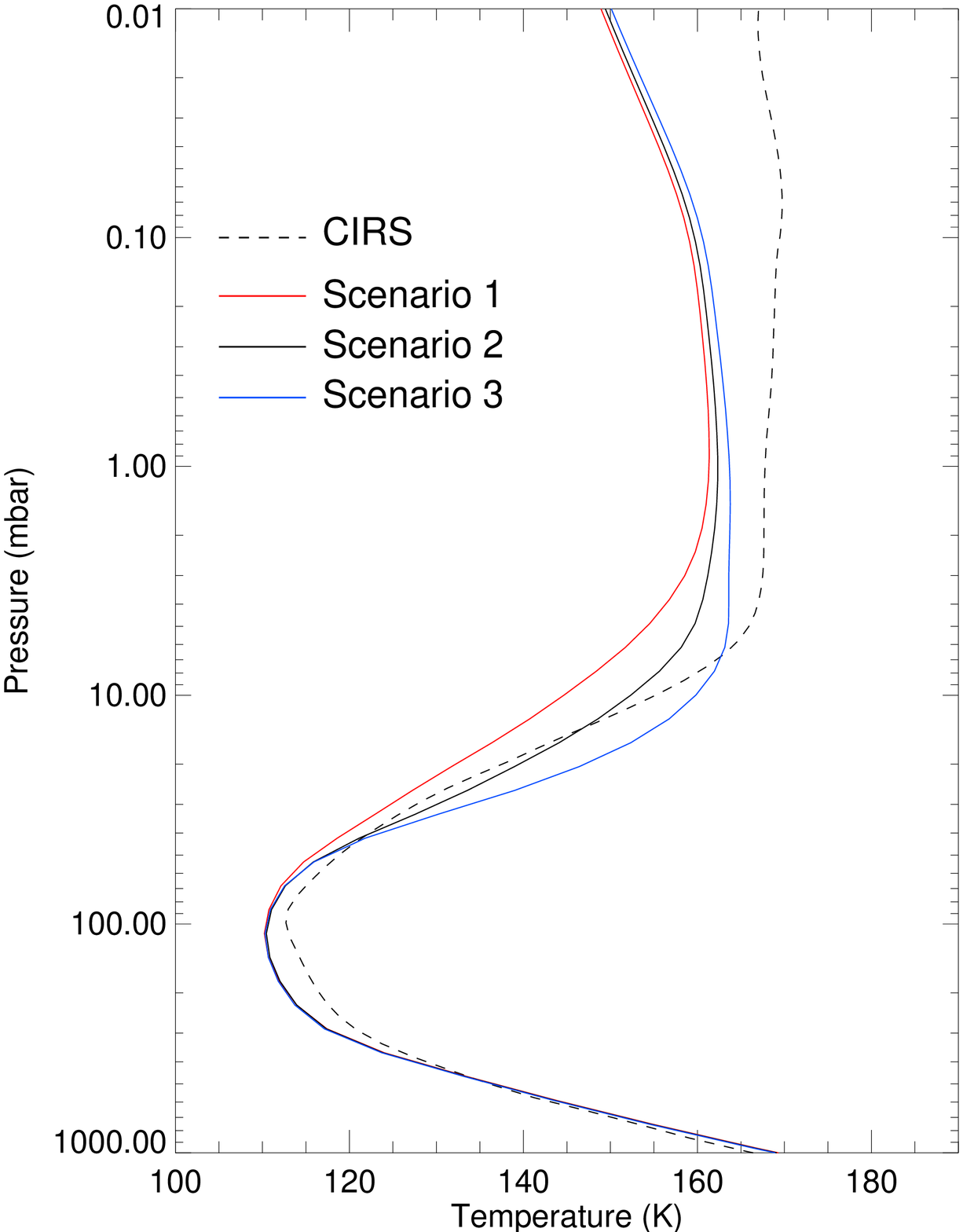}
  \caption{Comparison between temperature vertical profiles averaged between 77°S and 77°N as derived by \citet{Fletcher2016} from Cassini/CIRS observations, in dashed lines, and that predicted by our radiative equilibrium model, in solid lines, for the three polar haze scenario.}
  \label{fig:compa_avg_profiles}
\end{figure}

\begin{figure}
  \centering
  \includegraphics[width=0.7\linewidth]{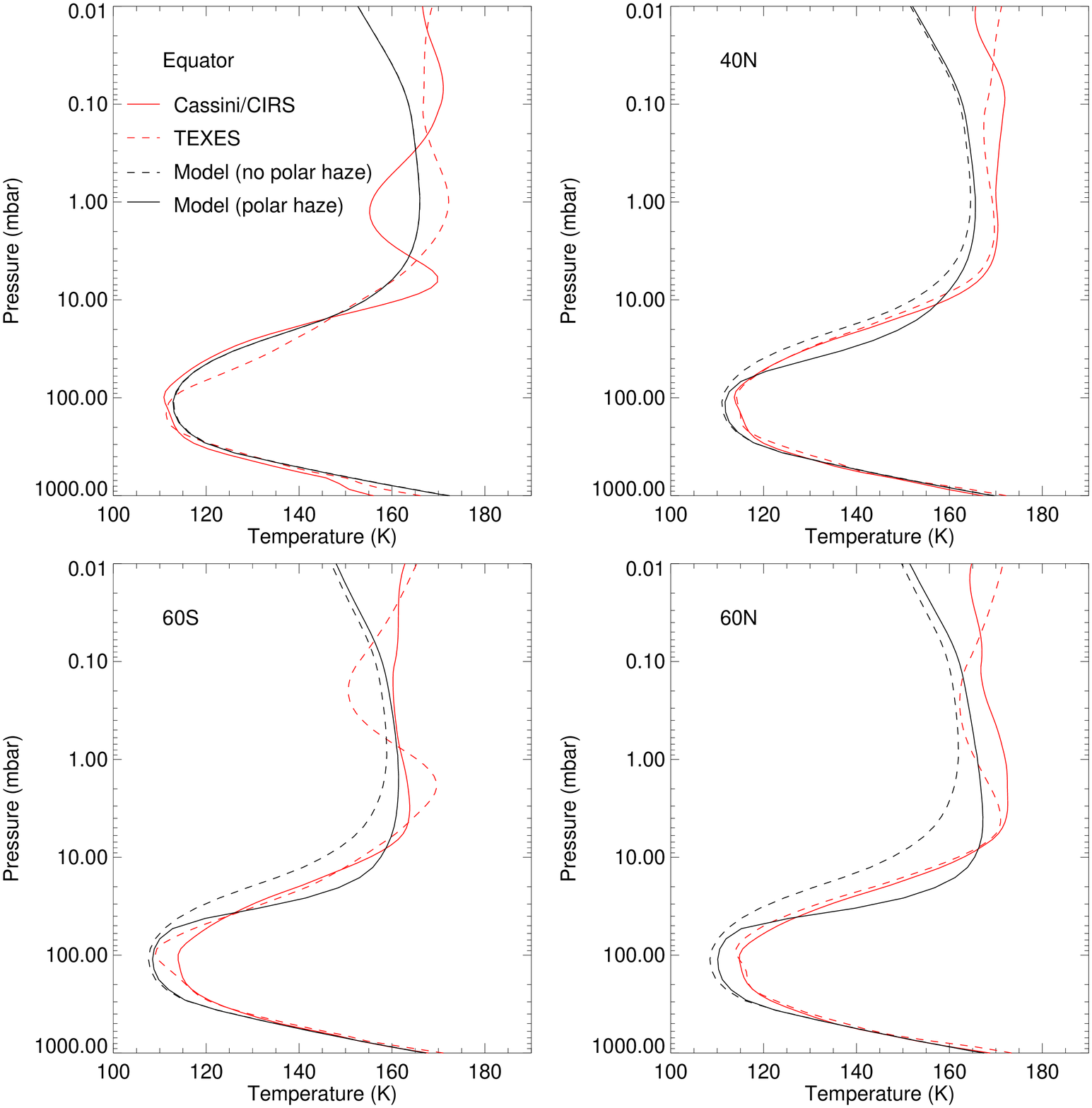}
  \caption{Comparison of vertical profiles of the temperature as derived from our radiative-convective model at four different latitudes (equator, 40\textdegree N, 60\textdegree S, 60\textdegree N), in solid black line, and observed temperature profiles derived from Cassini/CIRS (solid red line) and TEXES (dashed red line) derived by \citet{Fletcher2016}. Model results correspond to polar haze scenario~\#2. For reference, we also show the temperature predicted by our model without the polar haze (dashed black line).  }
  \label{fig:compa_fletcher_profiles}
\end{figure}

Regarding the stratosphere, we present in Figure~\ref{fig:compa_fletcher_strato} the meridional temperature variations at four pressure levels: 0.4, 3, 10 and 25~mbar. Results obtained with the three polar haze scenario are shown. 
Figure~\ref{fig:compa_avg_profiles} compares the modeled globally-averaged vertical profiles of temperature, for the three haze scenario, to that derived from CIRS observations.
Several individual vertical profiles of temperature at radiative equilibrium (this time, only for scenario~\#2 for the sake of clarity) are compared to CIRS and TEXES data at four latitudes in Figure~\ref{fig:compa_fletcher_profiles}. 
We first note that at all pressure levels, CIRS and TEXES exhibit a strong temporal variability at and near the equator. This region is known to harbor a periodic equatorial quasi-quadriennal oscillation (QQO) in the temperature and associated thermal wind field thought to result from wave-mean zonal flow interactions \citep{Leovy1991, Orton1991,Flasar2004,Simon2007}, based on analogy with similar oscillations on the Earth and Saturn. Hence, in the following, we will not comment on the model-observation mismatch near the equator, since by design our radiative-convective model cannot capture such a dynamical signature.

At the 10-mbar pressure level, our modeled temperatures exhibit a significant variability depending on the chosen polar haze scenario (see Figure~\ref{fig:compa_fletcher_strato}). We find that scenario~\#2 provides an excellent match to the observed temperatures. 
At this pressure levels, both CIRS and TEXES temperatures feature a local maximum at 50--65\textdegree N which is rather well reproduced by our model, should the polar haze scenario~\#2 be used. If the radiative impact of the polar haze is neglected, the temperature would be 10~to 15~K colder at these pressure levels and latitude range. 
With the polar haze scenario~\#1, the comparison with observations is even less favorable, as the haze has a net cooling effect at this pressure level. 
The hemispherical asymmetry  between latitudes 60\textdegree N  and 60\textdegree S, of about 8~K, observed by TEXES and CIRS, is faithfully reproduced as well with polar haze scenario~\#2 and ~\#3. We can argue that this observed hemispherical asymmetry is caused by a radiative effect related to the polar haze absorption, as this  asymmetry would be of only 1--2~K without this radiative contribution. As already discussed in section~\ref{sec:seasons} and shown in Figure~\ref{fig:tempe_seasons}, this asymmetry is a seasonal effect: it should disappear around Ls=230° and reverse at Ls=240--320° (the temperature at 60°S is expected to be 2~K warmer than that at 60°N at that season).
We note that in their analysis of Voyager observations, shortly after autumn equinox (Ls=190\textdegree ), \citet{Simon2006} found that the temperature was $\sim$6~K warmer at 50\textdegree N compared to 50\textdegree S (measurements only extended to 50° latitude). This is compliant with our results: indeed, due to the strong asymmetry in the polar haze as constrained by Zhang et al. (2013), where the integrated opacity is found to be about four times greater at 50°N compared to 50°S (see Figure~\ref{fig:aero_zhang}), our model predicts that the temperature at 50°N remains warmer than that at 50°S throughout the year - provided that the haze hemispheric asymmetry persists.

Our model predicts that a similar north-south asymmetry between 60\textdegree N  and 60\textdegree S is still present at the 25-mbar level. This is at odds with CIRS and TEXES observations, which are nearly symmetric about the equator at this pressure level. 
In addition, the haze scenarios~\#2 and 3 significantly overestimate the high latitude temperatures, especially in the northern hemisphere.
On the one hand, this could suggest that we overestimate the aerosol content at this pressure level in the northern hemisphere -- but not in the southern one, as our predicted temperature agrees better with CIRS and TEXES observations at 50--70\textdegree S. 
The actual vertical profile of aerosol opacity could be different from the one parameterized in eq.~\ref{eq:profile_aurora} and/or be different between 60\textdegree N and 60\textdegree S. In particular, it is important to note that \citet{Zhang2013} only constrained the shape of the vertical aerosol profile between 25°N and 75°S. For latitudes poleward of 25°N, they assumed a similar vertical shape than that at the corresponding southern latitude. Given the large sensitivity of the temperature to this polar haze, it is crucial that future observational studies better characterize its vertical profile in the northern hemisphere as well.
On the other hand, we also note that CIRS and TEXES measurements have a rather low sensitivity to the temperature in the 20 to 60~mbar range, so that it is possible that part of the observation-model mismatch at 25~mbar is also due to a larger uncertainty in the observations at this level.

At 10 and 25~mbar, both observations and model predict a sharp decrease in temperature between 65\textdegree \ latitude and the poles. The temperature drop in our model is sharper than the observed one, but this is not surprising: indeed, our model predicts a marked drop due to the sharp decrease of net heating rates at high latitudes. However, it is expected that such a strong temperature gradient would cause dynamical activity (e.g. thermally-direct circulations, baroclinic instability, \ldots) that would act to counteract this gradient. The study of the associated stratospheric circulation and/or mixing processes is left to a future study.

At lower pressure levels (p$<$ 3~mbar), we note that our predicted temperatures are almost systematically underestimated by $\sim$5~K compared to TEXES and CIRS observations, regardless of the chosen polar haze scenario.
This is well visible in Fig.~\ref{fig:compa_avg_profiles} on the globally-averaged temperature profiles.
This suggests that either a radiative ingredient is missing or not well estimated in our model, or that the temperature is governed by other processes, such as dynamical heating by gravity wave breaking or by eddies.
The same conclusion was reached by \citet{Zhang2013b}, who discussed these two hypothesis but did not favor one or the other. In a sequel paper, \citet{Zhang2015} find that the atmosphere at pressures $<$ 3~mbar seems to depart from radiative balance (with an excess of radiative cooling at global scale), but the authors emphasize that cooling and heating rate profiles still agree with each other within error bars. 
It is indeed plausible to reconcile the observed temperature profile (within their $\sim$2~K error bars) with our calculated equilibrium temperatures, should the amount of ethane and acetylene be reduced by $\sim$30\% (see sensitivity studies in section~\ref{sec_gas}). 
The typical 1-$\sigma$ error bar on the retrieved abundance of these hydrocarbons is on the order of 20\% \citep{Nixon2010}, which makes this scenario plausible, but at the cost of a greater uncertainty on ethane and acetylene mixing ratios than previously thought.
Hence, this topic is still an open question.

At 0.4~mbar, we note that TEXES and CIRS observations exhibit important temporal variability at high latitudes (poleward of 55\textdegree ). The observed temperature also increases between 60 and 78\textdegree \ latitude in both hemispheres, which is at odds with our simple radiative-convective model. Clearly, other processes must control the temperature at these altitudes.
One hypothesis is that the temperature may be influenced by the precipitation of high-energy particles that could warm the atmosphere at high latitudes through Joule heating \citep{Sinclair2017}.

We can also comment on the comparison with  \citet{Sinclair2017} who analyzed TEXES and CIRS data specifically focusing on the polar regions. They highlighted a strong local temperature maximum at 1~mbar in Jupiter's auroral oval, which was hypothetically attributed to either Joule heating or absorption by aerosols. In our radiative-convective simulations, the temperature maximum associated to aerosol absorption is obtained at 10--20~mbar (and not 1~mbar), where the peak concentration of polar haze is parametrized in our model. However, our simulated conditions are not that of the auroral oval itself: \citet{Sinclair2017} interpretation of strong aerosol heating at 1~mbar can hold if there is a local maximum of aerosol absorption at this level in the auroral oval, which remains to be assessed.

This comparison work with state-of-the-art observations shows that our radiative-convective equilibrium model with polar haze scenario~\#2 reproduces well, to first order, the observed temperature in the upper troposphere and lower stratosphere (p$>$5~mbar), except in the equatorial region (where there is well-known dynamical activity). 
Other processes might be at play in controlling the temperature in the middle and upper stratosphere, and in the troposphere (belt/zone activity), but the reason behind the systematic $\sim$5~K cold bias at low latitudes in the upper stratosphere is still largely unknown. 
These results are consistent with the work by~\cite{Zhang2015}, who found that the lower and mid stratosphere was near radiative equilibrium should a polar haze be included, similar to our haze scenario~\#2.
Using polar haze scenario~\#1 yields too cold temperatures at high latitudes, especially near 10~mbar, while scenario~\#3 results in systematically too warm temperatures at high latitudes in the 5--30~mbar pressure range. 
The impact of assuming different haze scenario is also well visible at global scale (see Fig.~\ref{fig:compa_avg_profiles}).
However, we caution that the objective of this comparison work is not to "fine tune" the haze properties until a match with observations is found, as a local radiative imbalance could trigger some atmospheric circulation in the actual atmosphere, modifying in turn the temperature.
Residual-mean circulations induced by such radiative imbalance are estimated in the following section.
In summary, while the extreme scenario 1 and 3 seem unlikely (given the magnitude of the observation-model mismatch and their systematic nature), our choosing scenario~\#2 does not rule out other possible combinations of haze properties. 
The best way forward is to 1) better characterize the polar haze in Jupiter's stratosphere while 2) exploring the stratospheric dynamics with the help of a global circulation model for different haze scenarios, and confront these future model results back to the observed temperatures.



\section{Residual-mean stratospheric circulations}
\label{circu}
In this section, we exploit the computed heating and cooling rates to estimate the residual-mean circulation in Jupiter's stratosphere.
We will in particular explore the impact of assuming different polar haze properties  on the residual-mean stratospheric circulation.

\subsection{Background}

Stratospheric circulations are driven by a combination of diabatic and mechanical (eddy-induced) forcings, resulting in an interplay of transport processes: advection and mixing. 
 In the Earth stratosphere, it has been shown that the Transformed Eulerian Mean (or residual-mean) circulation is a good approximation to the Lagrangian mean circulation (relevant to tracer transport) in regions where wave breaking and dissipation is relatively weak \citep{Dunkerton1978,Butc:14}. As we describe in what follows, the residual-mean circulation can be approximately estimated from the knowledge of atmospheric net heating rates and temperatures. Hereafter we follow this approach to diagnose the zonally-averaged mass circulation in Jupiter's stratosphere, for annually-averaged conditions. 

The complete equations for the zonally-averaged stratospheric circulation are provided by the Transformed Eulerian-mean formulation, with the respectively residual-mean meridional and vertical components of the circulation ~$(v^*,w^*)$ defined as a combination of a zonal-mean and eddy-induced terms \citep{andrews1987}:
\begin{equation} \label{eq:vstar}
v^*
=
\overline{v}
-\frac{1}{\rho_0} \,
\frac{\partial}{\partial{z}}
\left( \frac{\rho_0 \,\overline{v'\theta'}}{\partial \overline{\theta} / \partial z} \right)
\end{equation}
\begin{equation} \label{eq:wstar}
w^*=\overline{w}+\frac{1}{a\cos\phi}\frac{\partial}{\partial{\phi}}
\left(
\frac{\cos\phi \, \overline{v'\theta'}}{\partial \overline{\theta} / \partial z}
\right)
\end{equation}
\noindent where 
overlines denote zonal averages,
primes departures from the zonal mean (eddies),
$\theta$ potential temperature, 
$\rho_0$ density,
$a$ planetary radius,
$\phi$ latitude,
$z$ altitude.
The associated streamfunction~$\Psi$ describing the circulation is defined by
\begin{equation} \label{eq:streamf}
(v^*,w^*) = 
\frac{1}{\rho_0\cos\phi} \,
\left(
-\frac{\partial{\Psi}}{\partial{z}},
\frac{1}{a}\frac{\partial{\Psi}}{\partial{\phi}} 
\right)
\end{equation}
\noindent The two components~$(v^*,w^*)$ of the residual-mean circulation follow  a mass-conservation equation 
\begin{equation} \label{eq:cmass}
\frac{1}{a\cos\phi}\,
\frac{\partial \left( \cos\phi \, v^* \right)}{\partial\phi}
+\frac{1}{\rho_0} \, \frac{\partial \left( \rho_0 \, w^* \right) }{\partial{z}}=0
\end{equation}
\noindent and an energy-conservation (thermodynamic) equation
\begin{equation} \label{eq:cenergy}
\frac{\partial{\overline{\theta}}}{\partial{t}}+\frac{v^*}{a}\frac{\partial{\overline{\theta}}}{\partial{\phi}}+w^*\frac{\partial{\overline{\theta}}}{\partial{z}}=\mathcal{Q} + \mathcal{E}
\end{equation}
\noindent in which~$\mathcal{Q}$ is the net radiative heating rate and~$\mathcal{E}$ is the heating rate related to eddies forcing the mean flow
\begin{equation} \label{eq:eddies}
\mathcal{E} = - \frac{1}{\rho_0} \, \frac{\partial}{\partial{z}}\left[\rho_0
\left( \frac{\overline{v'\theta'} }{a}
\frac{\partial \overline{\theta} / \partial \phi}{\partial \overline{\theta} / \partial z}
+\overline{w'\theta'} \right) \right]
\end{equation}
For quasi-geostrophic large-scale flows in non-acceleration conditions, the eddy-related term~$\mathcal{E}$ can be neglected. Under this approximation, the residual-mean circulation is similar to the so-called diabatic circulation also used on Earth to diagnose the Brewer-Dobson circulation \citep{Butc:14}. Additionally, considering atmospheric state and circulation averaged over a year, the temporal term in eq.~\ref{eq:cenergy} can also be neglected, which entails
\begin{equation} \label{eq:cenergyapprox}
\frac{v^*}{a}\frac{\partial{\overline{\theta}}}{\partial{\phi}}+w^*\frac{\partial{\overline{\theta}}}{\partial{z}}\simeq\mathcal{Q}
\end{equation}
\noindent It is important to note here that neither the seasonal variations of temperature nor the impact of eddies are negligible  in Jupiter's stratosphere; yet the approximations are reasonable in a context where we seek the average meridional and vertical transport experienced by the long-lifetime chemical species in Jupiter's stratosphere. 

\subsection{Method}

Equations~\ref{eq:cmass} and~\ref{eq:cenergyapprox} are solved to obtain the residual-mean circulation~$(v^*,w^*)$ under the approximations~$\mathcal{E} \simeq 0$ (we neglect the eddy heat flux convergence term) and~$\partial \overline{\theta} / \partial t \simeq 0$ (we neglect seasonal variations). 
In equation~\ref{eq:cenergyapprox}, the temperature profiles~$\overline{\theta}(z)$ are an averaged of Cassini/CIRS and IRTF/TEXES observations analyzed by \citet{Fletcher2017}.
The choice of averaging these two data sets is motivated by the will to smooth out a bit the large amplitude of the QQO signal and to get a better representation of a seasonally averaged temperature at Ls$\sim$140°.
We compute the net radiative heating rates~$\mathcal{Q}$ by running our seasonal radiative model for just one time step, starting at Ls=140° and with a temperature field corresponding to the observed temperatures interpolated on our model grid. 
We repeat this work for the three polar haze scenario described in section~\ref{sec:aero_aurora} to test the sensitivity of the diabatic circulation to these different radiative forcings.

To solve equations~\ref{eq:cmass} and~\ref{eq:cenergyapprox}, we use the iterative method described in \citet{Solo:86} :
\begin{enumerate}
\item At the initial iteration~$i=0$, the meridional component~$v^*_{i=0}$ is set to zero in equation~\ref{eq:cenergyapprox}, and we simply solve for the vertical component~$w^*_{i=1}$ given the vertical gradient of potential temperature (as if simply computing adiabatic warming/cooling by subsiding/ascending motions equilibrating the radiative heating rate).
\item \label{recursive} The vertical component~$w^*_{i=1}$ is used to obtain the streamfunction~$\Psi_{i=1}$ by integrating equation~\ref{eq:streamf} (using a Simpson integration method)
\begin{equation} \label{eq:streamfint}
\Psi_{i=1} = \int_{-\frac{\pi}{2}}^{\phi} (w^*_{i=1}+\epsilon) \cos \phi \, a \, \textrm{d}\phi
\end{equation}
\noindent where 
\begin{equation} \label{eq:corr}
\epsilon = \left( \int_{-\frac{\pi}{2}}^{\frac{\pi}{2}} \cos \phi \, \textrm{d}\phi \right)^{-1} \int_{-\frac{\pi}{2}}^{\frac{\pi}{2}} w^*_{i=1} \cos \phi \, \textrm{d}\phi
\end{equation}
is a corrective term (usually a couple percent at best) designed to ensure that the streamfunction~$\Psi_i$ vanishes at the north pole. 
\item The meridional component~$v^*_{i=1}$ is obtained from the streamfunction~$\Psi_{i=1}$ by using equation~\ref{eq:streamf} (vertical derivative of~$\Psi_{i=1}$). Using the streamfunction~$\Psi_{i=1}$ to compute~$v^*_{i=1}$ from~$w^*_{i=1}$ is equivalent to solving the mass-conserving equation~\ref{eq:cmass}.
\item The meridional component~$v^*_{i=1}$ is injected in equation~\ref{eq:cenergyapprox} to obtain the vertical component~$w^*_{i=2}$ at the next iteration, then the process is looped back to step~\ref{recursive} for~$i=2$.
\end{enumerate}
\noindent This iterative procedure converges quickly: iterations~$i>3$ yield a change from~$(v^*_i,w^*_i)$ to~$(v^*_{i+1},w^*_{i+1})$ of about 1~\%. We stopped the computations at the tenth iteration in which the increment from the previous step is only 0.01~\%. 
Our algorithm 
was checked upon a well-constrained analytical example.

\subsection{Results and comparison to previous studies}
\label{res_circu}

\begin{figure}
  \centering
  \includegraphics[width=0.75\linewidth]{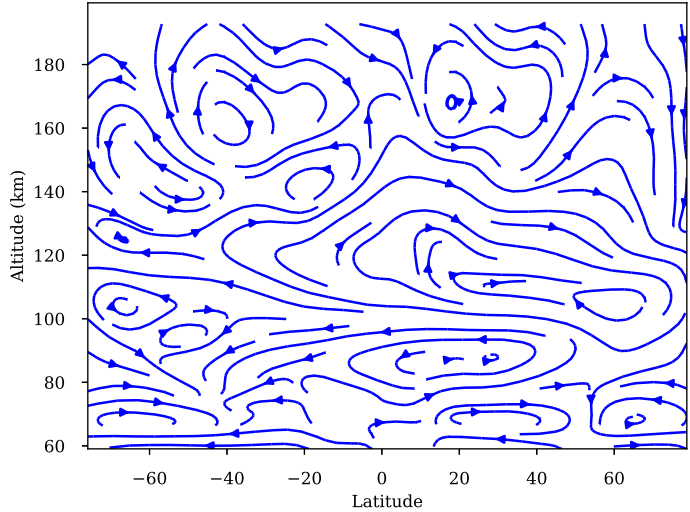}
  \caption{Streamlines computed from eq.~\ref{eq:streamfint}, using the polar haze scenario~\#2 and the averaged temperature derived from Cassini/CIRS and TEXES. The altitude  is computed with the convention z=0~km at the 1-bar level. For reference, the bottom of the figure, at 50~km, corresponds to the lower stratosphere ($\sim$50~mbar), while the 1~mbar level lies at $\sim$135~km. For the sake of clarity, the vertical component has been multiplied by 900 in this figure since the horizontal scale from one pole to the other, in km, is $\sim$900 times the vertical extent considered here (140~km).  }
  \label{fig:streamlines}
\end{figure}

\begin{figure}
  \centering
  \includegraphics[width=0.99\linewidth]{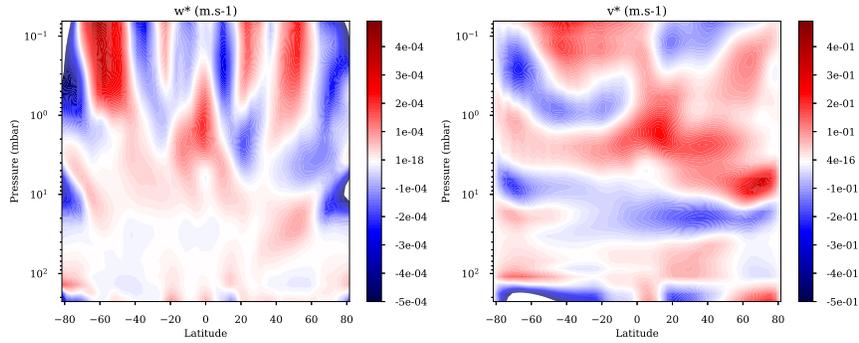}
  \caption{Pressure-latitude cross-section of the vertical (left) and meridional (right) components of the residual-mean circulation, in m.s$^{-1}$, estimated using the polar haze scenario~\#2 and temperatures derived from Cassini/CIRS and TEXES. }
  \label{fig:windspeed}
\end{figure}

\begin{figure}
  \centering
  \includegraphics[width=0.75\linewidth]{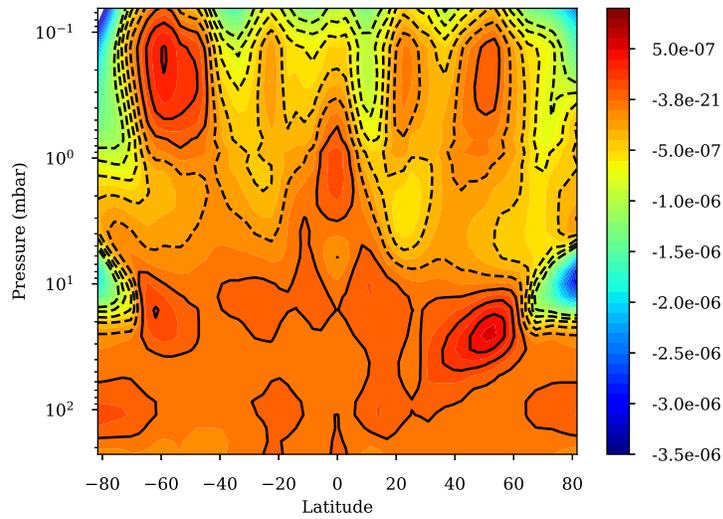}
  \caption{Pressure-latitude cross-section of the net heating rates in Jupiter's stratosphere, in K.s$^{-1}$, estimated using the polar haze scenario~\#2 and temperatures derived from Cassini/CIRS and TEXES.}
  \label{fig:net_q}
\end{figure}

Hereafter, we mostly describe the results obtained with the most favorable (according to section~\ref{sec:temperature}) polar haze scenario~\#2.
The resulting streamlines of the residual circulation, for the stratosphere only, are displayed in Figure~\ref{fig:streamlines} and the corresponding vertical and meridional wind speeds are shown in Figure~\ref{fig:windspeed}. For reference, the pressure-latitude cross section of the net radiative  heating rates~$\mathcal{Q}$ used to derive this circulation is shown in Figure~\ref{fig:net_q}.

\begin{figure}
  \centering
  \includegraphics[width=0.75\linewidth]{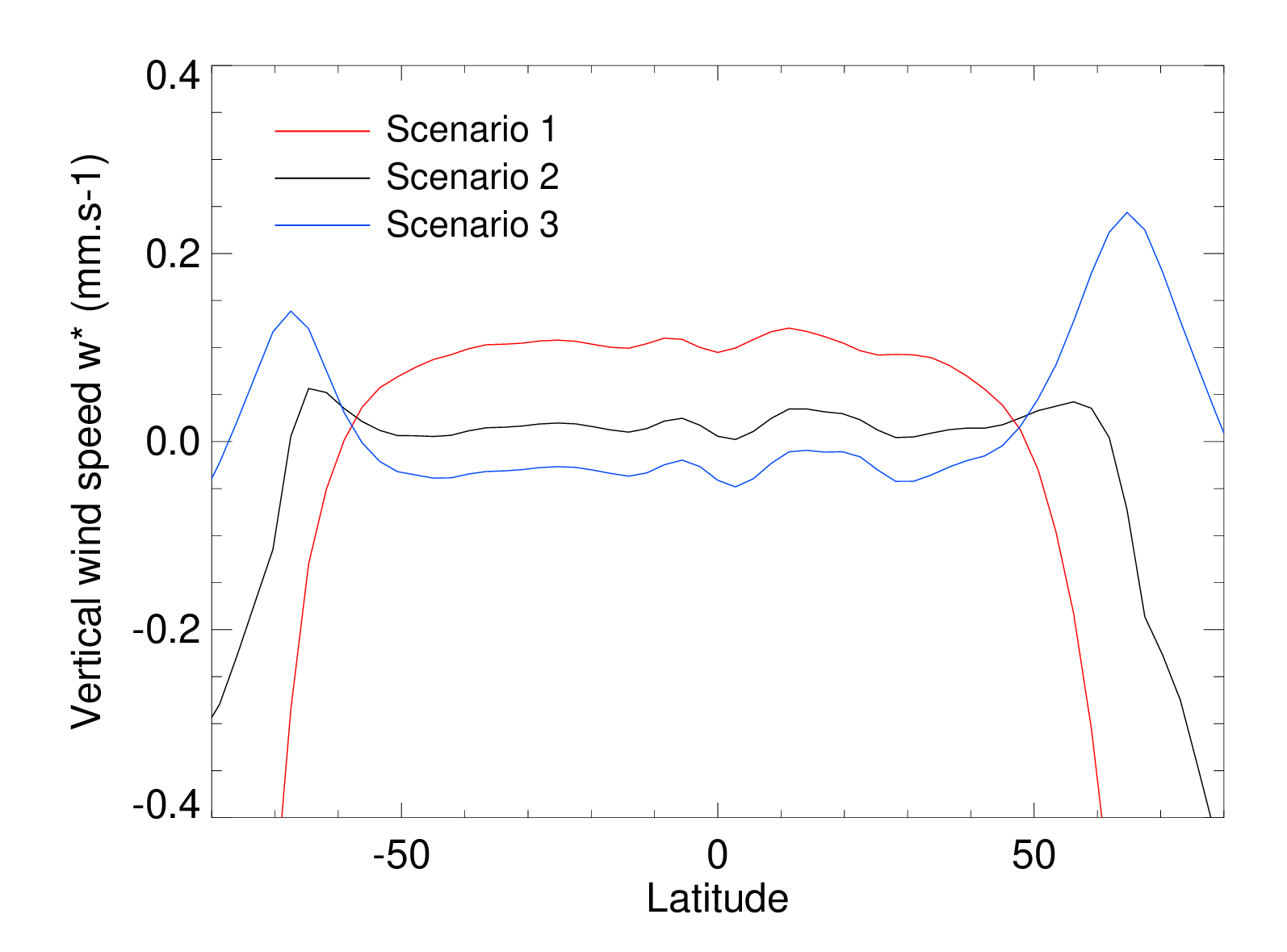}
  \caption{Vertical component of the residual-mean circulation at the 20-mbar pressure level, in mm.s$^{-1}$, estimated using the polar haze scenario~\#1, 2 or 3, as labeled.}
  \label{fig:compa_w}
\end{figure}

Overall, many small circulation cells are present, due to the fact that there are many local extrema in the spatial distribution of net heating rates~$\mathcal{Q}$. Nevertheless, two prominent large-scale circulation cells can be noted.
In the lower stratosphere (10 to 30~mbar, or $\sim$80--110~km altitude), the residual-mean circulation is characterized by upwelling at 50--60°N and cross-equatorial flow from northern high latitudes to southern mid latitudes. 
The vertical wind speed at 50°N reaches 0.12~mm.s$^{-1}$ at the 20-mbar level and the meridional wind speed is of the order of 0.10--0.15~m.s$^{-1}$. This cell is forced by the net positive heating rate centered at 50°N at a pressure level of 20~mbar. 
Indeed, as discussed in section~\ref{sec:compa_obs}, the expected equilibrium temperature at this location is much warmer than that observed by CIRS and TEXES. 
Subsequently, when using the observed temperatures to diagnose the circulation, an upwelling is needed to balance the "too cold" observed temperature. In other words, the net radiative heating is compensated by a diabatic cooling. 
Should the polar haze scenario~\#1 be used (characterized by net negative heating rate near 50--70°N, 3--20~mbar), the circulation would reverse, with a strong downwelling occurring at high latitudes near 5--10~mbar and equator-to-pole meridional wind centered at 5~mbar, in both hemispheres.
The sensitivity of the diabatic circulation to the polar haze scenario is illustrated in Figure~\ref{fig:compa_w}, which shows the vertical wind $w^*$ at the 10-mbar pressure level for each of the three haze scenario.
This illustrates well that accurate knowledge of radiative forcings is crucial to employ this method to understand the jovian stratospheric circulation.
By extension, this knowledge will also be crucial for future GCM simulations.

In the middle stratosphere ($p<$3~mbar or $z>$120~km), the circulation does not depend on the chosen polar haze scenario. The diabatic circulation there is dominated by upwelling in the 20°S--0° region with cross-equatorial meridional flow and subsidence poleward of 50°N. 
Vertical wind speed at the equator reaches 0.25~mm.s$^{-1}$ at the 2-mbar pressure level while the meridional wind speed in the northern hemisphere is of the order of 0.2--0.3~m.s$^{-1}$.
The significant upwelling motion near the equator and 1~mbar is clearly associated with the thermal structure of the Quasi-Quadriennal Oscillation (QQO) and its particular phase at the time of Cassini/CIRS observations in 2000 (see figure~\ref{fig:compa_fletcher_profiles}).
Averaging CIRS temperature profiles with the ones retrieved from TEXES smooths out partially this feature, but not entirely.
The observed temperature field exhibits a local minimum near 1~mbar (figure~\ref{fig:compa_fletcher_strato}), which translates into upwelling (hence, diabatic cooling) when diagnosing the circulation.
Given the periodic nature of the QQO signal, this feature is unlikely to be a part of the annually-averaged meridional circulation.

The fact that the dominant meridional motion is from equator to high northern latitudes, with stronger downwelling at high northern than at high southern latitudes, stems from a hemispheric asymmetry in the net heating rates. This asymmetry itself results from an asymmetry in the observed temperature at high latitudes (see figure~\ref{fig:compa_fletcher_strato}). 
However, it is unclear whether this asymmetry reflects an overall seasonal effect or more transient conditions. In addition, observed departures from the expected equilibrium temperature in polar regions might be caused by processes (Joule heating, etc) other than the dynamical motions that are diagnosed with this method. 
Another potential issue is related to the systematic underestimation, by our radiative equilibrium model, of the observed temperature by $\sim$5K at p$<$3~mbar, as discussed in section~\ref{sec:compa_obs}.
Energy transfer by eddies could be important at these pressure levels and is one possible explanation of this mismatch. If that is the case, then neglecting $\mathcal{E}$ in equation~\ref{eq:cenergy} is not valid and using the observed (warmer) temperatures for computing the cooling rates  while neglecting $\mathcal{E}$ when estimating the streamfunction could lead to erroneous results. 
Hence, this is another example of the limitations of this diabatic circulation method, which requires a high degree of confidence on both the observed temperature field and its drivers. 

Previous estimates of the residual-mean circulation in Jupiter's atmosphere were obtained  by \citet{West1992} and \citet{Moreno1997} using a similar formalism than ours, albeit based on the temperature field derived by Voyager/IRIS  to compute the cooling rates.
These temperature profiles had a low vertical resolution in the stratosphere and were provided at only two pressure levels: 77~mbar and 1~mbar. 
For intermediate pressure levels, \citet{West1992} and \citet{Moreno1997} had to rely on interpolation.
Hence, one strong limitation to their study is that the observational constraint they used (the Voyager/IRIS temperature fields) did not capture the  polar haze region, mainly located near 10--30~mbar. 
\citet{West1992} noticed that using this temperature field resulted in an imbalance between cooling and heating rates at global scale, with residual net radiative heating at the 10-mbar pressure level. 
While the authors did include the effect of absorption of UV and visible light by a stratospheric polar haze for the computation of solar heating rates, they most probably underestimated the cooling rates due to a lack of detailed knowledge of the thermal structure. 

\begin{figure}
  \centering
  \includegraphics[width=0.47\linewidth]{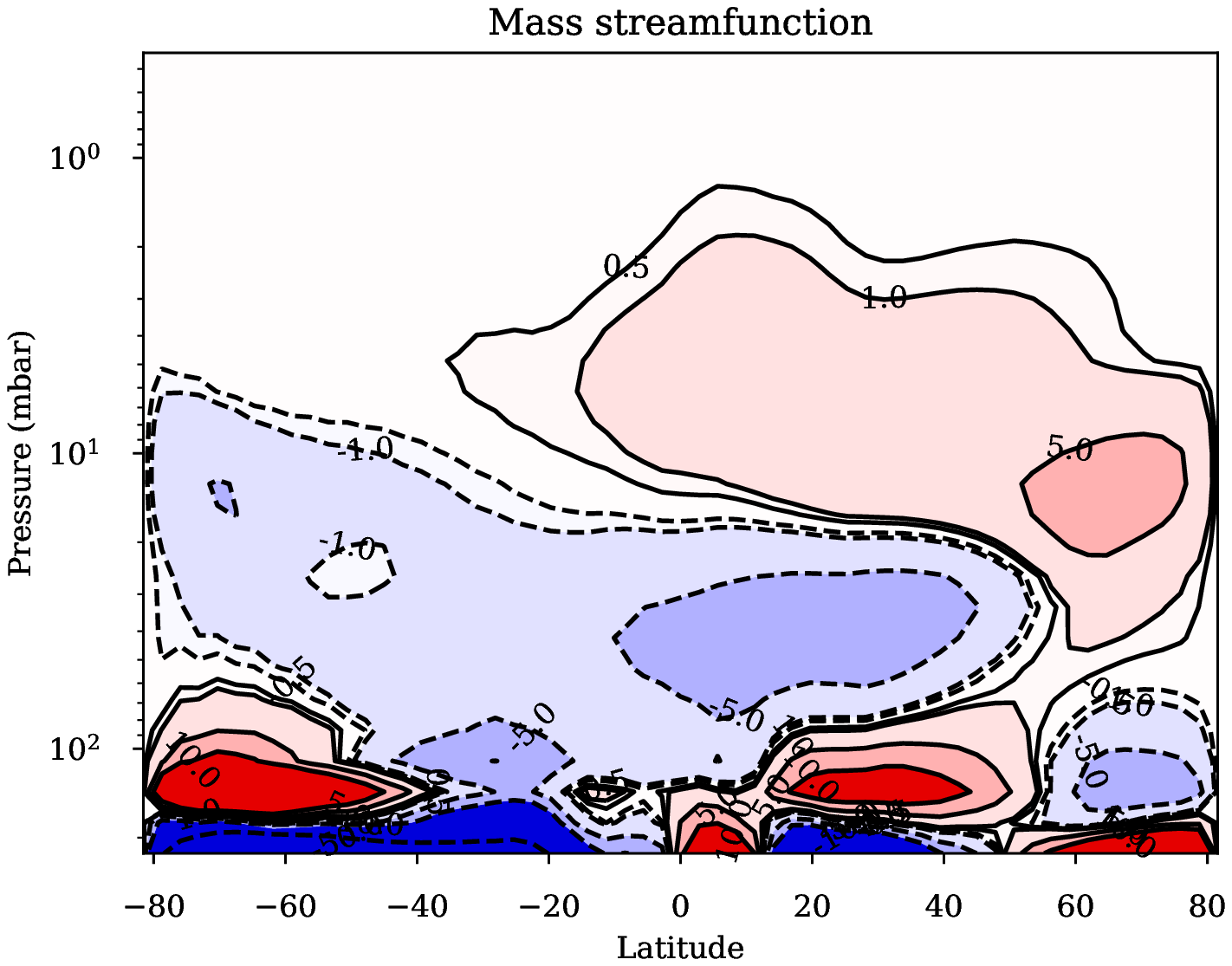}
  \includegraphics[width=0.47\linewidth]{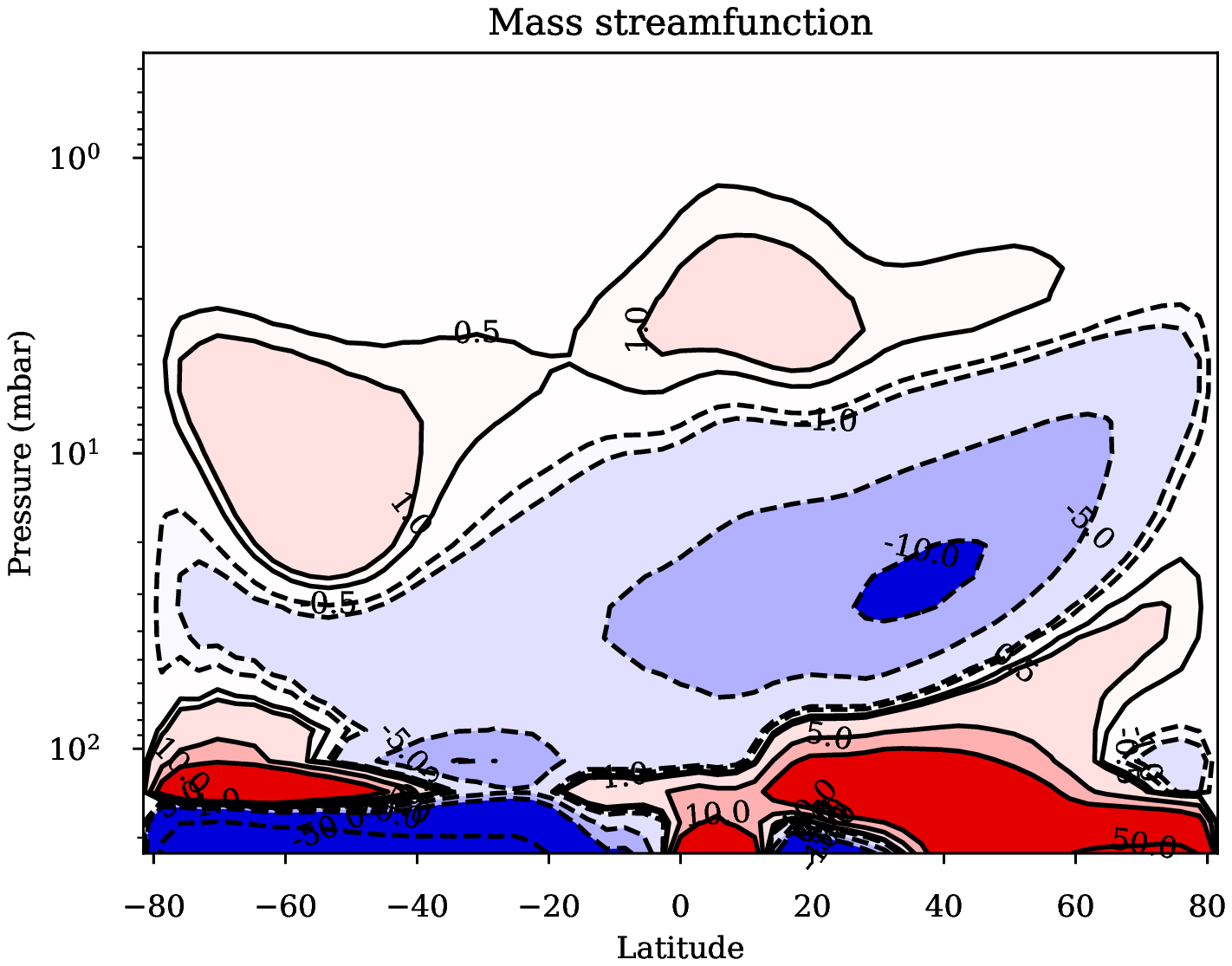}
  \caption{Mass streamfunction, in units of kg.m$^{-1}$.s$^{-1}$, computed from eq.~\ref{eq:streamfint} using either the polar haze scenario~\#2 (left panel) or scenario~\#3 (right panel). Results are shown up to 0.3~mbar to facilitate comparison with previous work by \cite{West1992} (note that the latter used different units, in g.m$^{-1}$.s$^{-1}$, so that a factor of 10 exists between the numerical values in this figure and fig.~7c of \citet{West1992}).}
  \label{fig:mass_stream}
\end{figure}

To mitigate this issue, \citet{West1992} scaled either the solar heating rate, or the infrared cooling rate, by a factor that depended on altitude only.
\citet{Moreno1997} noted a similar issue and chose to scale the temperature profile with a factor that depended on height but not with latitude, until radiative balance was achieved at global scale.
We can question these choices, as the resulting cooling rate would most certainly be missing a latitudinal-varying term (linked to the meridional distribution of the stratospheric haze and its impact on the thermal structure).
In addition, neither study included the contribution from the polar haze when computing the cooling rates; they only took into account gaseous contributions, which is now known to be largely inaccurate \citep{Zhang2015}.
Hence, for these reasons, the circulations derived by \citet{West1992} and \citet{Moreno1997} were probably flawed by these shortcomings.

Nevertheless, there are similarities with the circulation derived in this paper. The diabatic circulation estimated by \citet{West1992} is characterized by two cells with upwelling branches centered at 60°N and 70°S and at a pressure level of 3--10~mbar, driven by significant net radiative heating rates at these locations.
Maximum vertical wind speeds of the order of 10$^{-4}$~m.s$^{-1}$ (shown in Fig. 4 of \citet{Friedson1999}) are reached near 5~mbar.
The dominant meridional flow in \citet{West1992} is from high latitudes to low latitudes in the 3--30~mbar range, with subsidence occurring over a broad tropical region.
In our derived circulation with the reference scenario~\#2, an upwelling branch with similar vertical wind speed is present in the northern hemisphere, although located slightly deeper, near the 20-mbar level. Its counterpart in the southern hemisphere is muche weaker.
This difference with \citet{West1992} likely results from differences in the vertical distribution of haze and/or in the vertical temperature profile, as argued above.
However, considering the polar haze scenario~\#3 (with more absorbant haze), upwelling branches are present in both hemispheres, similarly to that derived by \citet{West1992}.
To illustrate this effect and facilitate the comparison with previously published work, we show in Figure~\ref{fig:mass_stream} the mass streamfunctions derived when assuming either scenario~\#2 or~\#3 that can be compared with Fig.~7(c) of \citet{West1992}.
The mass streamfunction obtained with scenario~\#3 is actually very similar to that derived by \citet{West1992}, both qualitatively and quantitatively.
This is not surprising: as discussed above, we suspect that \citet{West1992} analysis underestimated locally the actual cooling rates at the location where the polar haze warms the atmosphere. This resulted in large net heating rates at 60-70°S. Our scenario~\#3 exhibits large net heating rates as well, although for other reasons (due to greater absorption by the polar haze). 
This explains the similarity in the circulation derived in both these studies.
This represents another illustration of the sensitivity of the results to the assumed haze properties and to the observed temperature field used to compute cooling rates.


Finally, we can compare our derived residual vertical wind speeds with that of the eddy diffusion coefficient $K_{zz}$ taken from models A, B and C of \citet{Moses2005}.
In the range 1 to 30~mbar, estimates of $K_{zz}$ are of the order of 2$\times$10$^3$ to 10$^4$~cm$^2$.s$^{-1}$.
An order of magnitude of the vertical velocity induced by eddy motions can be obtained by dividing  $K_{zz}$ by the atmospheric scale height (25~km). This results in vertical wind speeds linked with eddy diffusivity of 1 to 4 $\times$10$^{-5}$~m.s$^{-1}$, \textit{ie.} just half the typical values of $w^*$ obtained in our study.


\subsection{Implications for the transport of trace species}

As discussed above, the circulation derived by our study -- even though using more detailed temperature fields and state-of-the-art opacities -- suffers from significant uncertainties as well. Given this uncertainty, we did not attempt to evaluate the impact of this circulation on the distribution of trace species.
However, we can comment on several existing studies, based on the lessons learned from this exercise.
For instance, \citet{Hue2018} study the distribution of ethane and acetylene, the main by-products of methane photochemistry,  with the goal of explaining the observed increase in ethane towards high latitudes, while acetylene is decreasing. The authors combine a photochemical model with a simple parametrization of transport (both advection and diffusion). In doing so, they test several Hadley-like circulation cells with upwelling at the equator and subsidence at both poles. 
The authors fail to explain the opposed distributions of acetylene and ethane.
Our study suggests that assuming such circulation cells is not appropriate in the lower stratosphere, where these puzzling hydrocarbon distributions are observed. 
More complex circulation patterns, such as those shown in Figure~\ref{fig:mass_stream}, are probably at play and need to be further understood.

Another puzzle is related to CO$_2$ and HCN, products of the SL-9 impact that occurred at 44\textdegree S,  which display opposite trends several years after the impact (CO$_2$ being maximum at the south pole while HCN is found well-mixed from mid-southern to mid-northern latitudes).
One hypothesis proposed by \citet{Lellouch2006} is that HCN and CO$_2$ were deposited at different altitudes and were transported by different wind regimes.
If we assume that the diabatic circulation derived from the polar haze scenario~\#2 is realistic, then this scenario implies that HCN was deposited near 0.5--5~mbar, to be transported equatorward, and CO$_2$ was deposited at pressures either lower than 0.5~mbar or around 10~mbar to be transported poleward  (see Figure~\ref{fig:windspeed}). Such scenario need to be tested in the future with chemistry-transport models.

\section{Conclusions}
\label{discussion}

We have developed a radiative-convective equilibrium model for Jupiter's troposphere and stratosphere that includes parametrizations of several cloud and haze layers. As for its Saturn counterpart \citep{Guerlet2014}, this model is computationally efficient and aims at being coupled with a dynamical core of a General Circulation Model \citep[as was recently done in the Saturn case, see ][]{Spiga2020}.
We take into account the radiative contribution of :

\begin{itemize}
    \item CH$_4$, C$_2$H$_6$, C$_2$H$_2$ and NH$_3$ for radiatively active species along with collision-induced absorption from H$_2$--H$_2$, H$_2$--He;
    \item A rather compact ammonia cloud located at 840~mbar comprising 10-$\mu$m particles, with a visible integrated opacity of 10.
    \item A tropospheric haze layer extending between 660 and 180~mbar composed of 0.5 $\mu$m particles with near "grey" optical constants and an integrated opacity of 4.
    \item A stratospheric haze layer made of fractal aggregates (typically, 1000 monomers of 10-nm each, fractal dimension of 2) supposedly linked with precipitation of high-energy particles at high latitudes. Their opacity is maximum near the 20-mbar level and at latitudes poleward of 30\textdegree N and 45\textdegree S.
\end{itemize}

The gaseous abundance profiles as well as tropospheric cloud and haze layer properties are fixed in latitude and time in the current version of our model, but we have studied the sensitivity to varying those parameters. For instance, varying the abundance of hydrocarbons with latitude in similar proportions than the observed ones (i.e. a poleward enhancement in ethane by a factor of two, while acetylene is reduced by 50\%) lead to temperature changes of at most 4~K in the 1 to 0.1-mbar level. Increasing the cloud opacity by a factor of two yields a temperature increase of 3~K in the upper troposphere. These changes are rather small and our main conclusions are not hampered by these simplistic assumptions. The inclusion of photochemistry (that would compute realistic hydrocarbon variations) or cloud microphysics (that would simulate spatio-temporal evolution of cloud formation on Jupiter's atmosphere) is devoted to a future study.

We confirm that the stratospheric polar aerosols have an important role in the radiative budget of Jupiter's stratosphere, yet with a significant uncertainty regarding its magnitude. 
Their net impact at latitudes 45--60\textdegree \ is maximum in the range 5--30~mbar and depends on their assumed properties (refractive index, size and number of monomers).  
A large contribution of aerosols to the heating rates was already demonstrated by \citet{Zhang2015} \citep[and before that by][]{West1992}, but it is the first time that the impact of aerosols on stratospheric temperatures is studied.
We tested the response of the atmosphere (in terms of radiative equilibrium temperatures) to three different polar haze scenario and find that the reference model of \citet{Zhang2013} provides a satisfactory comparison to observations at the time of the Cassini flyby in 2000 (Ls=110\textdegree ) or to a TEXES observing run in 2014 (Ls=176\textdegree ). The other two models tend to either systematically overestimate or underestimate the observed temperature in the 10--30~mbar range.

We find that the hemispheric asymmetry in stratospheric aerosols opacity (that is much larger in the northern than in the southern hemisphere) combined with the small obliquity and eccentricity  of Jupiter cause the predicted 10-mbar temperature to be systematically warmer at 50\textdegree N than at 50\textdegree S, throughout the year.
The asymmetry of 8~K in temperature at the 10-mbar level reported by \citet{Fletcher2016} between 60°N and 60°S  is also well reproduced by our model with polar haze scenario~\#2.
We also find that the polar haze significantly shortens the radiative timescales, estimated in section~\ref{sec:rad_times} to 100 days, or 3\% of a Jupiter year, at the 10-mbar level.
Nevertheless, significant uncertainties remain regarding the optical properties, sizes, meridional distribution and temporal variations  of this stratospheric polar haze. This prevents an advanced interpretation of the model-observations mismatch.

At lower pressures ($p<$3~mbar), we find that the modeled temperature is systematically lower than the observed one, by typically 5~K.
This is consistent with the previous study of radiative budgets in giant planet atmospheres by \citet{ChengLi2018}, who find that the cooling rate excesses the heating rate in a large part of Jupiter's stratosphere, and with previous studies by \citet{Zhang2013b} and \citet{Kuroda2014}.
However, as already noted by \citet{Zhang2015}, radiative equilibrium could be reached considering the  uncertainties on heating and cooling rates associated with the uncertainty on the abundance of hydrocarbons.
In our case, warmer equilibrium temperatures can be reached if we assume that ethane and acetylene are currently overestimated by roughly 30\%.
Other possible explanations are: a missing radiative ingredient ; a mechanical forcing (such as gravity wave breaking or other eddy terms warming the atmosphere) ; a coupling with thermospheric or ionospheric circulations, through Joule heating, adiabatic compression or horizontal advection \citep[e.g. ][ although the upper stratosphere is maginally covered in their study]{Majeed2005}. 
These scenario need to be further evaluated.

In theory, and under the assumption that the eddy heat flux convergence term is negligible, the knowledge of net radiative heating rates can be then exploited to estimate the stratospheric residual-mean circulation, in a similar fashion as \citet{West1992}.
In the Earth's stratosphere, the residual-mean circulation represents well, on a seasonal scale, the transport of tracers in regions where wave breaking and dissipation are weak \citep[][and references therein]{Butc:14}. This topic is of high interest on Jupiter, as the observed meridional distribution of photochemical products (ethane and acetylene) or by-products of comet Shoemaker Levy 9 impact (HCN, CO$_2$, dust...) is puzzling and cannot be explained by simple chemistry-transport models.
We revisited the study by \citet{West1992} and \citet{Moreno1997} by estimating the diabatic circulation based on state-of-the-art knowledge on opacity sources and observed temperature fields from the Cassini fly-by of Jupiter in December 2000 and ground-based TEXES observations in 2014.
Our main conclusion is that our current limited knowledge on the different radiative forcing terms (in particular regarding the stratospheric haze properties) and mechanical forcings (related to the magnitude of eddy heat flux) results in a low-to-moderate confidence in the estimated circulation.
On Earth, both the temperature field and the net radiative heating rates are known with a much higher degree of confidence, allowing to derive the main circulation patterns from this method.
The lessons learned from trying to adapt this method to Jupiter are that more investigations are needed regarding the characterization of Jupiter's polar haze radiative properties and other drivers of the meridional circulation.
Observations of the polar regions are challenging from Earth, but feasible. The Juno spacecraft offers unique views of the poles and could help characterizing the haze in the UV and near infrared. The future JUICE mission will also include a specific science phase at relatively high obliquity to get good views of Jupiter's polar regions, from the UV to 5$\mu$m. 

 To conclude, we have documented here the building and validation of a radiative-convective model for Jupiter, discussed the resulting equilibrium temperature, how it compares with observations and attempted to derive the residual-mean circulation associated with the computed net heating rates. 
 In order to go further into understanding Jupiter's atmospheric circulations, both in the troposphere and stratosphere, current efforts are focused on running 3D GCM simulations for Jupiter using a hydrodynamical solver coupled with the radiative seasonal model described herein. This will give insights into -- among other topics -- understanding what is the role of eddies in controlling the stratospheric circulations, mixing and thermal structure ; what governs the distribution of trace species and what are the mechanisms driving the QQO. 
 These topics are also valid for Saturn's atmosphere, which shares similar open questions regarding its atmospheric circulation but with different seasonal forcings, opening the way to comparative studies between these two gas giants.

\section*{Acknowledgements}

S. Guerlet and A. Spiga  acknowledge funding by the French Agence Nationale de la Recherche (ANR) under grant agreements ANR-12-PDOC-0013, ANR-14-CE23-0010-01 and ANR-17-CE31-0007.
Part of this work was funded by CNES as a support for Cassini/CIRS data interpretation.
We thank Xi Zhang (UC Santa Cruz) for providing their retrieved aerosol number density map, Jeremy Burgalat and Pascal Rannou (Reims University) for sharing their library to generate optical constants for fractal aggregates and Michael Rey (Reims University) for providing linelists for methane and its isotopologues (available through the TheoReTS plateform: http://theorets.univ-reims.fr/).
We thank the two anonymous referees for their very helpful and constructive comments that undoubtedly improved the quality of this manuscript.

\section*{Data availability}

The 3-D (latitude, pressure, time) temperature field for the reference radiative-convective simulation will be made available, in the form of a NetCDF file with a doi, on the data service hosted by Institut Pierre Simon Laplace when the paper will be accepted. 

\bibliographystyle{apalike} 
\bibliography{jupiter}

\end{document}